\documentclass[12pt]{article}
\usepackage[letterpaper, left=.9in, top=.8in, right=.9in, bottom=.4in,nohead,includefoot,verbose, ignoremp]{geometry}
\pdfoutput=1

\usepackage{booktabs}
\usepackage{natbib}
\usepackage{times} 
\usepackage{latexsym,epsfig,amssymb,amsmath,amsfonts,graphicx,amsthm,ifthen,pifont,comment,enumerate,setspace,multirow,color,footnote}
\usepackage{amscd, xy}
\input{xy}
\xyoption{all}

\def\begmat{\left(\begin{array}}\def\endmat{\end{array}\right)}

\def\bi{\begin{itemize}\setlength{\itemsep}{0pt}} \def\ei{\end{itemize}}

\def\bl{\begin{list}{\labelitemi}{\leftmargin=1em}\setlength{\itemsep}{-2.5pt}}  \def\el{\end{list}}
\def\bn{\begin{enumerate}} \def\en{\end{enumerate}}
\def\bt{\begin{table}[h]} \def\et{\end{table}}
\def\bc{\begin{center}} \def\ec{\end{center}}
\def\T{{ \mathrm{\scriptscriptstyle T} }}

\newcommand{\bfb} {\mbox{\boldmath $\beta$}}

\newcommand{\bfy} {\mbox{\boldmath $y$}}

\newcommand{\norm}[1]{\left\Vert#1\right\Vert}

\newtheorem{theorem}{Theorem}[section]
\newtheorem{lemma}[theorem]{Lemma}

\newtheorem{remark}[theorem]{Remark}

\newcommand \bbP{\mathbb{P}}
\newcommand \bbE{\mathbb{E}}

\def\m{\mathcal}

\def\mbx{\mbox}

\def\l{\left}
\def\r{\right}

\theoremstyle{plain}

\theoremstyle{plain}

\theoremstyle{remark}

\theoremstyle{plain}
\newtheorem{ass}{Assumption}[section]

\begin{document}


\title{Bayesian factorizations of big sparse tensors}

\author{Jing Zhou, Anirban Bhattacharya, Amy Herring and David Dunson}

\maketitle

\begin{center}
\textbf{Abstract}
\end{center}
It has become routine  to collect data that are structured as multiway arrays (tensors).  
There is an enormous literature on low rank and sparse matrix factorizations, but limited 
consideration of extensions to the tensor case in statistics.  The most common low rank tensor factorization
relies on parallel factor analysis (PARAFAC), which expresses a rank $k$ tensor as a sum of rank one tensors.
When observations are only available for a tiny subset of the cells of a big tensor, the 
low rank assumption is not sufficient and PARAFAC has poor performance.  We induce an additional layer
of dimension reduction by allowing the effective rank to vary across dimensions of the table.
For concreteness, we focus on a contingency table application.  Taking a Bayesian approach, we place priors 
on terms in the factorization and develop an efficient Gibbs sampler for posterior computation.  Theory is 
provided showing posterior concentration rates in high-dimensional settings, and the methods are shown to have excellent performance in 
simulations and several real data applications.
\vspace*{.3in}

\noindent\textsc{Keywords}: {Big data; Bayesian; Categorical data; Contingency table; Low rank; Matrix completion; PARAFAC; Tensor factorization.}

\section{Introduction}

Sparsely observed big tabular data sets are commonly collected in many applied domains.  One example corresponds to recommender 
systems in which the dimensions of the table correspond to users, items and different contexts (\cite{Karatzoglou2010}), with a tiny proportion of the cells filled in for users providing rankings. The task is to fill in the rest of the huge table in order to make recommendations to users of which items they may prefer in each context.  This extends the widely studied matrix completion problem (\cite{candesrecht2009}) of which the Netflix challenge was one example.  Another setting corresponds to contingency tables in which multivariate categorical data are collected for each individual, and the cells of the table contain counts of the number of individuals having a particular combination of values.  In contingency table analyses, the focus is typically on inferring associations among the different variables, but challenges arise when there are many variables, so that the number of cells in the table is vastly bigger than the sample size.

Suppose that the tensor of interest is $\pi \in \Pi_{d_1 \times \cdots \times d_p}$, with $\Pi_{d_1 \times \cdots \times d_p}$ a space of $p$-way tensors having $d_j$ rows in the $j$th direction.  Often there are constraints on the elements of the tensor.  For recommender systems, ratings are non-negative so that one is faced with a non-negative tensor factorization problem (\cite{paatero1994positive,lee1999learning,friedlander2005,limcomon09,liu2012}).  For contingency tables, the tensor corresponds to the joint probability mass function for multivariate categorical data, so that the elements are non-negative and add to one across all the cells (\cite{dunsonxing08,anirban2012}).  
Let $Y$ denote the data collected on tensor $\pi$.  
For recommender systems, $Y$ consists of ratings for a small subset of the $\prod_{j=1}^p d_j$ cells in the tensor, while for contingency tables $Y$ includes response vectors $y_i = (y_{i1},\ldots, y_{ip})^{\T}$ for subjects $i=1,\ldots,n$, with $y_{ij} \in \{1,\ldots,d_j\}$ for $j=1,\ldots,p$.  In both cases, data are extremely sparse, with no observations in the overwhelming majority of cells.

To combat this data sparsity, it is necessary to substantially reduce dimensionality in estimating $\pi$.  The usual way to accomplish this is through a low rank assumption.  Unlike for matrices, there is no unique definition of rank but the most common convention is to define the rank $k$ of a tensor $\pi$ as the smallest value of $k$ such that $\pi$ can be expressed as 
\begin{eqnarray}
\pi = \sum_{h=1}^k \psi_h^{(1)} \otimes \cdots \otimes \psi_h^{(p)}, \label{eq:parafac}
\end{eqnarray}
which is sum of $k$ rank one tensors, each an outer product of vectors\footnote{For $p = 2$, $\psi^{(1)} \otimes \psi^{(2)} = \psi^{(1)} \psi^{(2) \T}$. In general, $( \psi^{(1)} \otimes \cdots \otimes \psi^{(p)})_{c_1\ldots c_p} = \psi^{(1)}_{c_1} \ldots \psi_{c_p}^{(p)}$} for each dimension \citep{kolda09tensor}.  Expression (1) is commonly referred to as parallel factor analysis (PARAFAC) (\cite{harshman70,bro1997}).  For $k$ small, the number of parameters is massively reduced from $\prod_{j=1}^p d_j$ to $k \sum_{j=1}^p d_j$; as the low rank assumption often holds approximately, this leads to an effective approach in many applications, and a rich variety of algorithms are available for estimation.  

However, the decrease in degrees of freedom from exponential in $p$ to linear in $p$ is not sufficient when $p$ is big.  Large $p$ small $n$ problems arise routinely, and a usual solution outside of tensor settings is to incorporate sparsity.  For example, in linear regression, many of the coefficients are set to zero, while in estimation of large covariance matrices, sparse factor models are used that assume few factors and many zeros in the factor loadings matrices (\cite{West03, Carvalho08}).  In the matrix factorization literature, there has been consideration of low rank plus sparse decompositions (\cite{chartrand2012}), but this approach does not solve our problem of too many parameters.  Including zeros in the component vectors $\{ \psi_h^{(j)} \}$ is not a viable solution, particularly as we do not want to enforce exact zeros in blocks of the tensor $\pi$ but require an alternative notion of sparsity.

Our notion is as follows.  For component $h$ ($h=1,\ldots,k$), we partition the dimensions into two mutually exclusive subsets  $S_h \cup S_h^c = \{1,\ldots,p\}$.  The proposed sparse PARAFAC (sp-PARAFAC) factorization is then 
\begin{eqnarray}\label{eq:spPARAFAC0}
\pi = \sum_{h=1}^k \psi_h^{(1)} \otimes \cdots \otimes \psi_h^{(p)},\quad \psi_h^{(j)} = \psi_0^{(j)}\ \mbox{for $j \in S_h^c$}.
\end{eqnarray}
Hence, instead of having to introduce a separate vector $\psi_h^{(j)}$ for every $h$ and $j$, we allow there to be more degrees of freedom used to characterize the tensor structure in certain directions than in others.  Consider the recommender systems application and suppose we have three dimensions, including users ($j=1$), items ($j=2$) and context ($j=3$).  If we let $\psi_h^{(3)} = \psi_0^{(3)}$ for $h=1,\ldots,k$, 
\begin{eqnarray}
\pi_{c_1c_2c_3} = \psi_{0c_3}^{(3)} \sum_{h=1}^k \psi_{hc_1}^{(1)}\psi_{hc_2}^{(2)}, \label{eq:RSexample}
\end{eqnarray}
so that we factorize the user-item matrix as being of rank $k$, and then include a multiplier specific to each level of the context factor.  This assumes that users rank systematically higher or lower depending on context but there is no interaction.  In the contingency table application, $\mbox{Pr}( y_{i1}=c_1,\ldots, y_{ip}=c_p ) = \pi_{c_1\cdots c_p}$.  If $j \in S_h^c$ for $h=1,\ldots,k$, then the $j$th variable is independent of the other variables with $\mbox{Pr}( y_{ij} = c_j ) = \psi_{0c_j}^{(j)}$.  By including $j \in S_h^c$ for some but not all $h \in \{1,\ldots,k\}$ one can use fewer degrees of freedom in characterizing the interaction between the $j$th factor and the other factors.  In practice, we will learn $\{ S_h \}$ using a Bayesian approach, as the appropriate lower dimensional structure is typically not known in advance.

We conjecture that many tensor data sets can be concisely represented via (\ref{eq:spPARAFAC0}), with results substantially improved over usual PARAFAC factorizations due to the second layer of dimension reduction.  For concreteness and brevity, we focus on contingency tables, but the methods are easily modified to other settings.  Contingency table analysis is routine in practice; refer to \cite{agresti2002categorical,fienberg2007three}.  However, in stark contrast to the well developed literature on linear regression and covariance matrix estimation in big data settings, very few flexible methods are scalable beyond small tables. Throughout the rest of the paper, we assume that the observed data $y_i = (y_{i1}, \ldots, y_{ip})^{\T}, i = 1, \ldots, n$, is multivariate unordered categorical, with $y_{ij} \in \{1, \ldots, d_j \}$. Our interest is in situations where the dimensionality $p$ is comparable or even larger than the number of samples $n$. 


\section{Sparse Factor Models for Tables}

\subsection{\bf{Model and prior}}

We focus on a Bayesian implementation of sp-PARAFAC in (\ref{eq:spPARAFAC0}). Let $\m S^{r-1} = \{ x \in \Re^r: x_j \geq 0, \sum_{j=1}^r x_j = 1 \}$ denote the $(r-1)$-dimensional probability simplex.
In the contingency table case, \citet{dunsonxing08} proposed the following probabilistic PARAFAC factorization.
\begin{eqnarray}\label{eq:dx}
\mbox{Pr}( y_{i1}=c_1,\ldots, y_{ip}=c_p) = \pi_{c_1\cdots c_p} = \sum_{h=1}^k \nu_h \prod_{j=1}^p \lambda_{hc_j}^{(j)}, 
\end{eqnarray}
where $\nu = \{ \nu_h \} \in \m S^{k-1}$ and $\lambda_h^{(j)} = (\lambda_{h1}^{(j)}, \ldots, \lambda_{hd_j}^{(j)}) \in \m S^{d_j - 1}$ is a vector of probabilities of $y_{ij}  = 1, \ldots, d_j$ in component $h$. Introducing a latent sub-population index $z_i \in \{ 1, \ldots, k \}$ for subject $i$, the elements of $y_i$ are conditionally independent given $z_i$ with $\mbox{Pr}(y_{ij} = c_j \mid z_i=h) = \lambda_{hc_j}^{(j)}$, and marginalizing out the latent index $z_i$ leads to a mixture of product multinomial distribution for $y_i$.  Placing Dirichlet priors on the component vectors leads to a simple and efficient Gibbs sampler for posterior computation. We will refer to this model (\ref{eq:dx}) as standard PARAFAC. 

This approach has excellent performance in small to moderate $p$ problems, but as $p$ increases there is an inevitable breakdown point.  The number of parameters increases linearly in $p$, as for other PARAFAC factorizations, so problems arise as $p$ approaches the order of $n$ or $p \gg n$.  For example, we are particularly motivated by epidemiology studies collecting many categorical predictors, such as occupation type, demographic variables, and single nucleotide polymorphisms.  For continuous response vectors $y_i \in \Re^p$, there is a well developed literature on Gaussian sparse factor models that are adept at accommodating $p \gg n$ data (\cite{West03, Lucasetal06, Carvalho08, bhattacharya2011sparse}).  These models include many zeros in the loadings matrices to induce additional dimension reduction on top of the low rank assumption. \cite{debdeep2013} provided theoretical support  through characterizing posterior concentration.

Our sp-PARAFAC factorization provides an analog of sparse factor models in the tensor setting.  Modifying for the categorical data case, we let
\begin{align}\label{eq:spdx}    
\pi_{c_1 \ldots c_p} = \sum_{h=1}^k \nu_h \prod_{j \in S_h} \lambda_{hc_j}^{(j)} \prod_{j \in S_h^c} \lambda_{0c_j}^{(j)}, 
\end{align}
where $|S_h| \ll p$ ($|S|$ denotes the cardinality of a set $S$) and the $\lambda_0^{(j)}$ vectors are {\em fixed in advance}; we consider two cases: 
$$(i)\ \lambda_0^{(j)} = \bigg( \frac{1}{d_j},\ldots,\frac{1}{d_j} \bigg)^{\T}\quad \mbox{and}\quad
(ii)\ \lambda_0^{(j)} = \bigg( \frac{1}{n} \sum_{i=1}^n y_{i1},\ldots, \frac{1}{n} \sum_{i=1}^n y_{ip} \bigg)^{\T},$$ 
corresponding to a discrete uniform and empirical estimates of the marginal category probabilities. By fixing the baseline dictionary vectors $\{ \lambda_0^{(j)} \}$ in advance, and allocating a large subset of the variables within each cluster $h$ to the baseline component, we dramatically reduce the size of the model space.  In particular, the probability tensor $\pi$ in \eqref{eq:spdx} can be parameterized as 
$
\theta_{\pi} = \l(\nu, \{S_h\}_{1\leq h \leq k}, \{ \lambda_h^{(j)} \}_{1\leq h \leq k, j \in S_h} \r),
$
where $\nu \in \m S^{k-1}, S_h \subset \{1, \ldots, p\}, \lambda_h^{(j)} \in \m S^{d_j-1}$. 
Thus, the effective number of model parameters is now reduced to $(k-1) + \sum_{h=1}^k |S_h| + \sum_{h=1}^k \sum_{j \in S_h}  (d_j-1)$, which is substantially smaller than the $(k-1) + \sum_{j=1}^p k(d_j-1)$ parameters in the original specification, provided $|S_h| \ll p$ for all $h = 1, \ldots k$. This is ensured via a sparsity favoring prior on $|S_h|$ below. We will illustrate that this can lead to huge differences in practical performance.

Completing a Bayesian specification with priors for the unknown parameter vectors and expressing the model in hierarchical form, we have\footnote{$\mbox{Mult}\big(\{1, \ldots, d\}; \lambda_1, \ldots, \lambda_d \big)$ denotes a discrete distribution on $\{1, \ldots, d\}$ with probabilities $\lambda_1, \ldots, \lambda_d$ associated to each atom.}
\begin{eqnarray}
	& y_{ij} \sim \mbox{Mult} \big( \{1,\ldots,d_j\}; \lambda_{z_i1}^{(j)},\ldots,\lambda_{z_id_j}^{(j)}\big), \nonumber \\ 
	& \lambda_h^{(j)} \sim (1-\tau_h)\delta_{\lambda_0^{(j)}} + \tau_h \mbox{Diri} (a_{j1},\ldots,a_{jd_j}), \nonumber \\
	& \mbox{Pr}(z_i = h) =  \nu_h = V_h \prod_{l<h}(1-V_l),\nonumber \\
	& V_h \sim \mbox{Beta}(1,\alpha), \quad \alpha \sim \mbox{Gamma} (a_{\alpha}, b_{\alpha}), \quad \tau_h \sim \mbox{Beta}(1,\gamma). \label{eq:spPARAFAC}
\end{eqnarray}
It is evident that the hierarchical prior in \eqref{eq:spPARAFAC} is supported on the space of probability tensors with a sp-PARAFAC decomposition as in \eqref{eq:spdx}, since \eqref{eq:spPARAFAC} is equivalent to letting the subset-size $|S_h| \sim \mbox{Binom}(p, \tau_h)$ and drawing a random subset $S_h$ uniformly from all subsets of $\{1, \ldots, p\}$ of size $|S_h|$ in \eqref{eq:spdx}. 
A stick-breaking prior \citep{sethuraman1994constructive} is chosen for the component weights $\{ \nu_h \}$, taking a nonparametric Bayes approach that allows $k=\infty$, with a hyperprior placed on the concentration parameter $\alpha$ in the stick-breaking process to allow the data to inform more strongly about the component weights.  The probability of allocation $\tau_h$ to the {\em active} (non-baseline) category in component $h$ is chosen as $\mbox{beta}(1, \gamma)$, with $\gamma>1$ favoring allocation of many of the $\lambda_h^{(j)}$s to the baseline category $\lambda_0^{(j)}$. In the limiting case as $\gamma \to \infty$, the joint probability tensor $\pi$ becomes an outer product of the baseline probabilities for the individual variables,
$
\pi = \lambda_{0}^{(1)} \otimes \cdots \otimes \lambda_{0}^{(p)}.
$
On the other hand, as $\gamma \to 0$, one reduces back to standard PARAFAC (\ref{eq:dx}). 

Line 2 of expression (\ref{eq:spPARAFAC}) is key in inducing the second level of dimensionality reduction in our Bayesian sparse PARAFAC factorization.  The inclusion of the baseline component that does not vary with $h$ massively reduces the number of parameters, and can additionally be argued to have minimal impact on the flexibility of the specification.  The $\lambda_h^{(j)}$s are incorporated within 
$\prod_{j=1}^p \lambda_{h c_j}^{(j)}$, which for large $p$ is highly concentrated around its mean since the $\lambda_h^{(j)}$'s are independent across $j$.  This is a manifestation of the concentration of measure phenomenon \citep{talagrand1996new}, which roughly states that a random variable that depends in a smooth way on the influence of many independent variables, but not too much on any one of them, is essentially constant.  For example, if $\theta_j \stackrel{iid}{\sim} U(0,1)$ and 
$\Theta = \prod_{j=1}^p \theta_j$, then $\mbox{E}( \Theta ) = (1/2)^p$ and $\mbox{var}( \Theta ) = (1/3)^p$, which rapidly converges to zero.  This implies that replacing a large randomly chosen subset of the $\lambda_h^{(j)}$s by $\lambda_0^{(j)}$ should have minimal impact on modeling flexibility.

\subsection{\bf{Induced prior in log-linear parameterization}}

An important challenge is accommodating higher order interactions, which play an important role in many applications (e.g., genetics), but are typically assumed to equal zero for tractability.  As $p$ grows, it is challenging to even accommodate two-way interactions in traditional categorical data models (log-linear, logistic regression) due to an explosion in the number of terms.  In contrast, the tensor factorization does not explicitly parameterize interactions, but indirectly induces a shrinkage prior on the terms in a saturated log-linear model.  One can then reparameterize in terms of the log-linear model in conducting inferences in a post model-fitting step.  We illustrate the induced priors on the main effects and interactions below.  

For ease of exposition, we first focus on a case where $p = 3$ and $d_j = d = 2$ for $j = 1, \ldots, 3$. We generate 
$10,000$ random probability tensors $\pi^{(t)} = ( \pi_{c_1 c_2 c_3}^{(t)}), t = 1, \ldots, 10,000$ distributed according to \eqref{eq:spPARAFAC}, where we fix the baseline $\lambda_0^{(j)} = (1/2, 1/2)$ for all $j$. Given a $2 \times 2 \times 2$ tensor $\pi$, we can equivalently characterize $\pi$ in terms of its log-linear parameterization 
$$
\bfb = (\beta_1, \beta_2, \beta_3, \beta_{12}, \beta_{13}, \beta_{23}, \beta_{123})^{\T},
$$
consisting of $3$ main effect terms $\beta_1, \beta_2, \beta_3$, three second-order interaction terms $\beta_{12}, \beta_{13}, \beta_{23}$ and one third order interaction term $\beta_{123}$; refer to \S 5.3.5 of \cite{agresti2002categorical}. Given each prior sample $\pi^{(t)}$, we equivalently obtain a sample $\bfb^{(t)}$ from the induced prior on $\bfb$, which allows us to estimate the marginal densities of the main effects and interactions and also their joint distributions. In particular, since $\gamma$ plays an important role in placing weights on the baseline component, we would like to see how our induced priors differ with different $\gamma$ values. 

\begin{figure}[htbp]
	\centering	\includegraphics[width=6.5in]{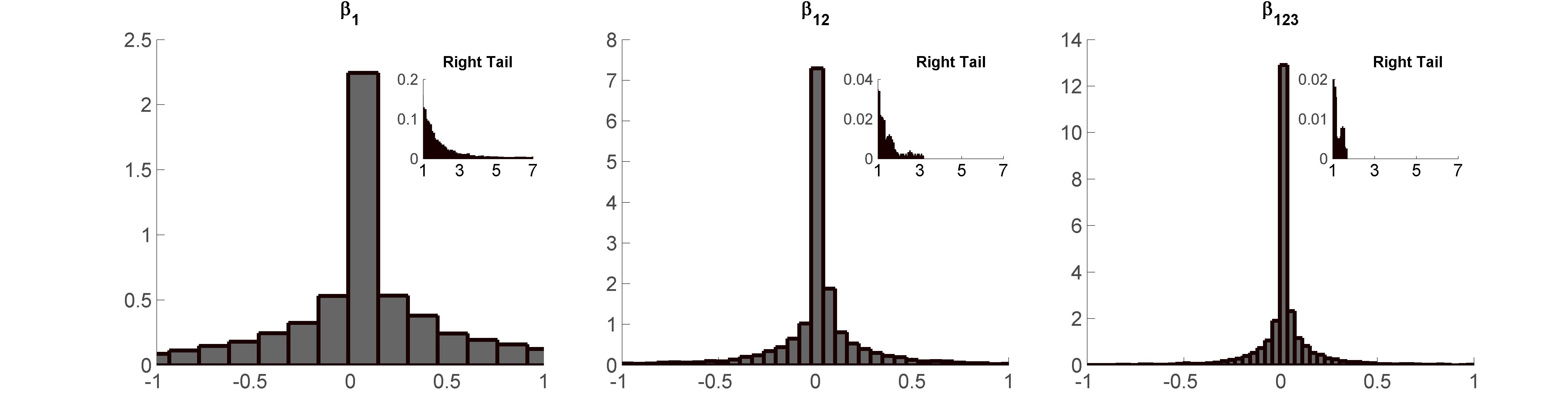}
	\includegraphics[width=6.5in]{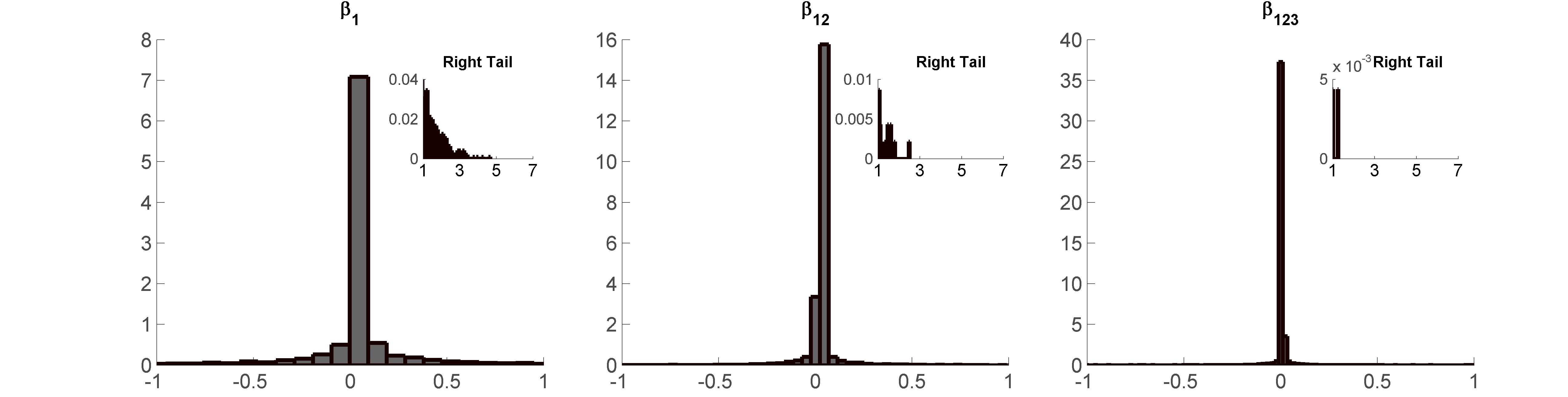}
	\includegraphics[width=6.5in]{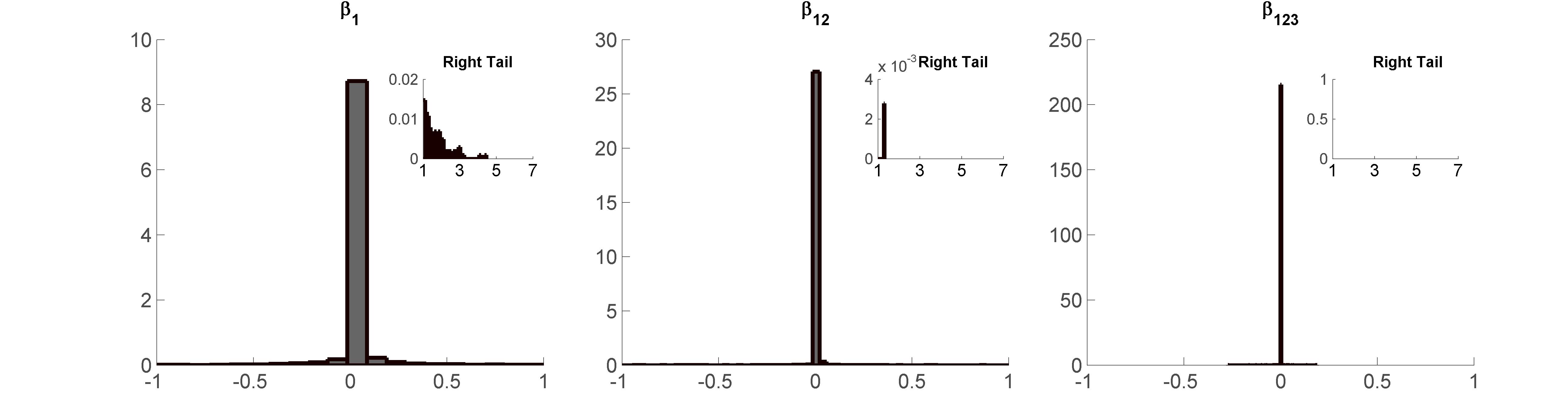}
		\caption{Histograms of induced priors for one main effect $\beta_1$, one two-way interaction $\beta_{12}$, and the three-way interaction $\beta_{123}$ - Top Row: $\gamma=1$;  Middle Row: $\gamma=5$;  Bottom Row: $\gamma=20$.}
	\label{fig:gamm1520}
\end{figure}

In our simulation exercise, we fix three values of $\gamma$, namely, $\gamma  = 1, 5, 20$. Note that $\gamma = 1$ corresponds to a $U(0, 1)$ prior on $\tau_h$. For different values of $\gamma$, we show the histograms of one main effect term $\beta_1$, one two-way interaction $\beta_{12}$ and the three-way interaction $\beta_{123}$ in Figure~\ref{fig:gamm1520}. Table~\ref{tab:gammtab} additionally reports summary statistics.

In high-dimensional regression, $y_i = x_i^{\T} \bfb + \epsilon_i$, there has been substantial interest in shrinkage priors, which draw $\beta_j$ {\em a priori} from a density concentrated at zero with heavy tails.  Such priors strongly shrink the small coefficients to zero, while limiting shrinkage of the larger signals \citep{park2008bayesian,carvalho2010horseshoe,polson2010shrink, hans2011elastic,armagan2011generalized}.  In Figure~\ref{fig:gamm1520},  the induced prior on any of the log-linear model parameters is symmetric about zero, with a large spike very close to zero, and heavy tails. Thus, we have indirectly induced a continuous shrinkage prior on the main effects and interactions through our tensor decomposition approach.  In addition, the prior automatically shrinks more aggressively as the interaction order increases.  Such greater shrinkage of interactions is commonly recommended \citep{Gelman2008}.  Importantly, we do not zero out small interactions but allow many small coefficients, which is an important distinction in applications, such as genomics, having many small signals.  The hyperparameter $\gamma$ serves as a penalty controlling the degree of shrinkage.

Our next set of simulations involve larger values of $p$, where the necessity of the regularization implied by $\gamma$ becomes strikingly evident. In the log-linear parameterization, we now have $p$ main effects $\beta_1, \ldots, \beta_p$; let $\bfb_{main} = (\beta_1, \ldots, \beta_p)^{\T}$. In the $p \gg n$ setting, one cannot even hope to consistently recover all the main effects unless a large fraction of the $\beta_j$'s are zero or close to zero. One would thus favor a shrinkage prior on $\bfb_{main}$, with any particular draw resembling a near-sparse vector. Since the induced prior on the $\beta_j$'s is continuous, we study the $l_1$ norm $\norm{ \bfb_{main} }_1 = \sum_{j=1}^p | \beta_j|$ as a surrogate for the $l_0$ norm to quantify the sparsity. 

\begin{figure}[htbp]
\centering	\includegraphics[width=6.5in]{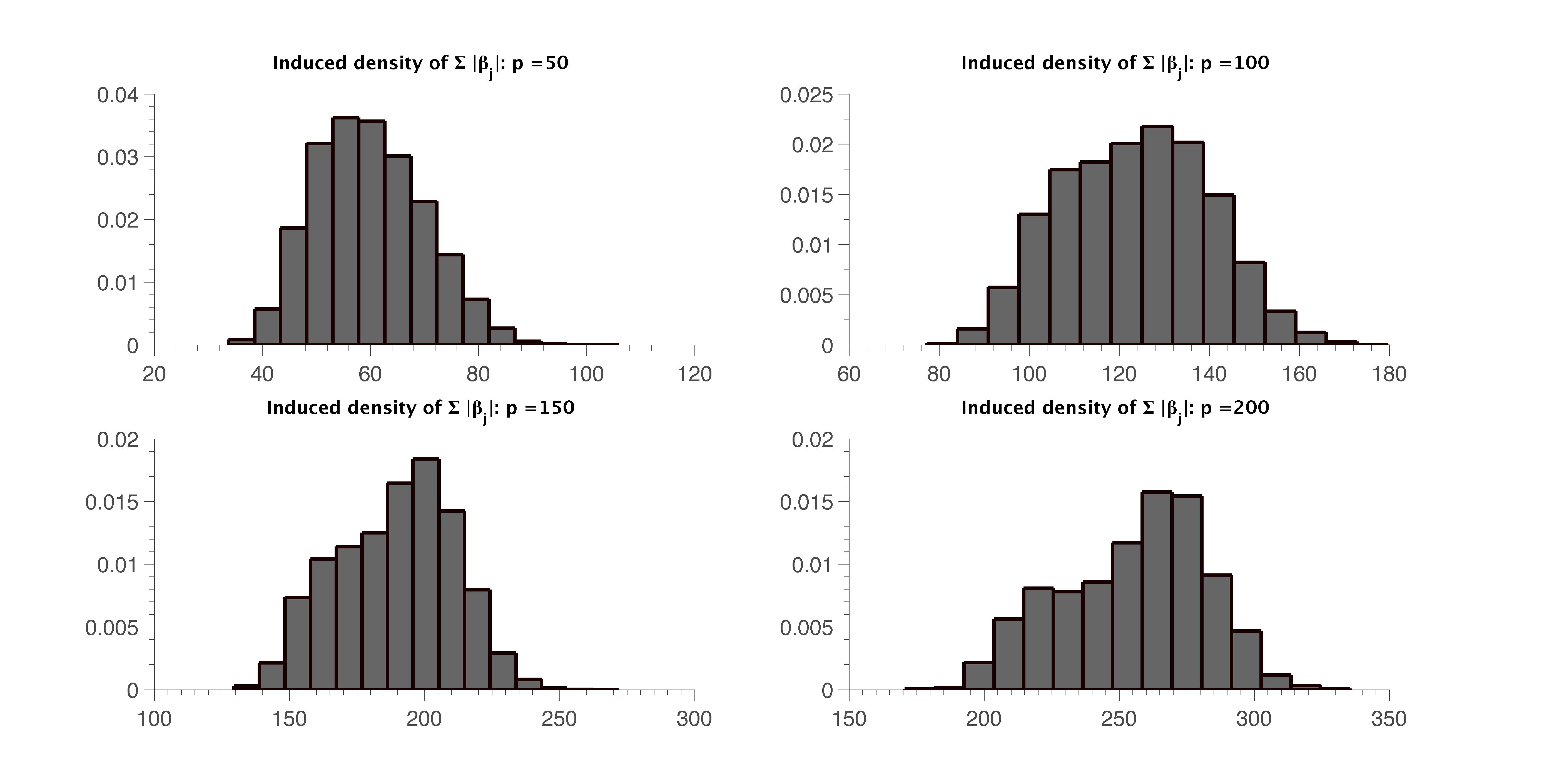}
\caption{Histograms of $\norm{ \bfb_{main} }_1$ for different values of $p$ under the standard PARAFAC model.}
\label{fig:gam_0}
\end{figure}

\begin{figure}[htbp]
\centering	\includegraphics[width=6.5in]{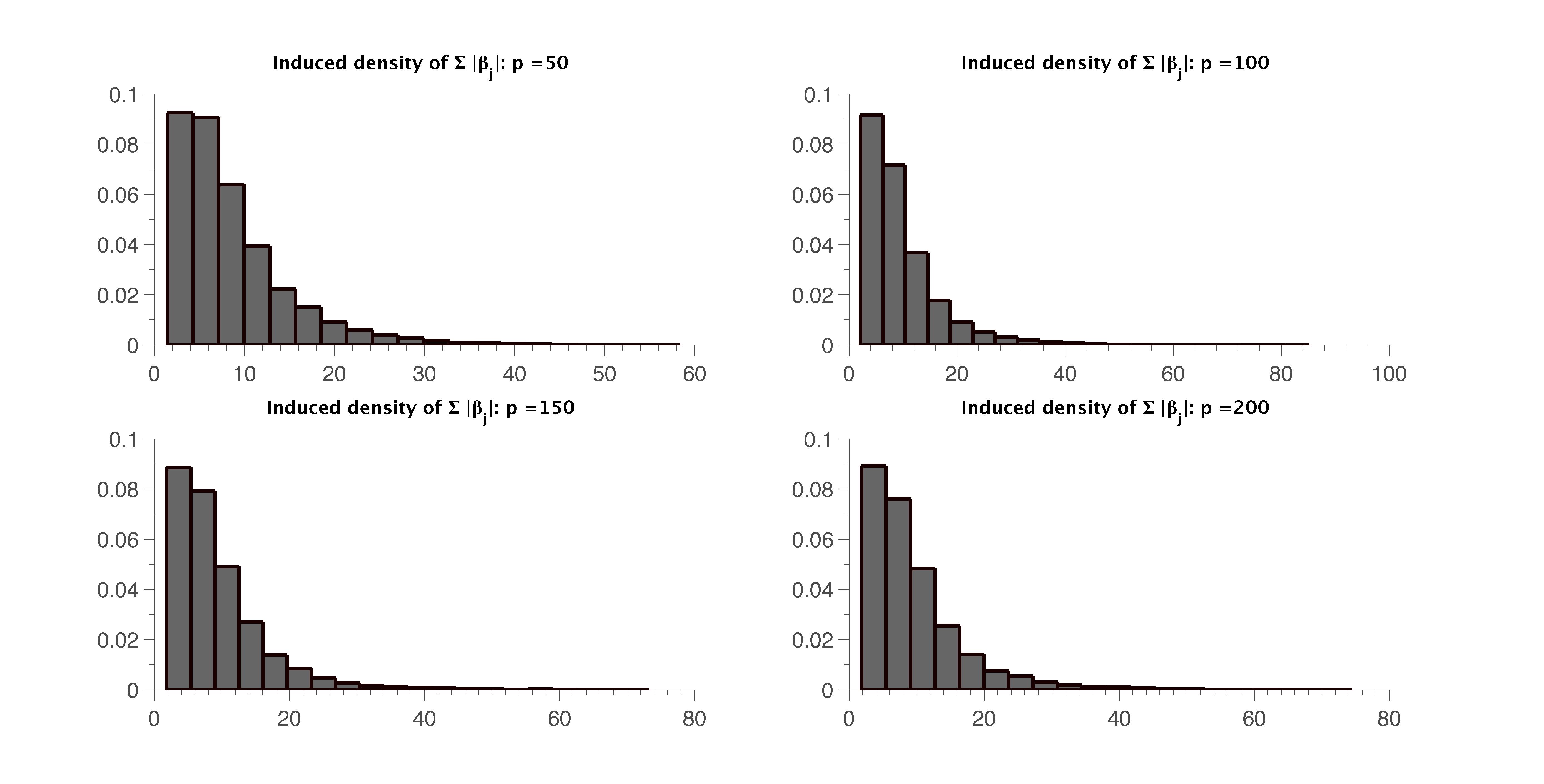}
\caption{Histograms of $\norm{ \bfb_{main} }_1$ for different values of $p$ under the sp-PARAFAC model with $\gamma = 0.1 p$.}
\label{fig:gam_var}
\end{figure}

We consider $p = 50, 100, 150, 200$ and plot histograms of the induced density of $ \norm{ \bfb_{main} }_1$ based on $10,000$ prior draws in Figures \ref{fig:gam_0} and \ref{fig:gam_var}. Figure \ref{fig:gam_0} corresponds to the case where $\gamma = 0$, i.e., when the sp-PARAFAC formulation reduces back to the standard PARAFAC \eqref{eq:dx}, while $\gamma/p$ is set to a constant $\beta \in(0,1)$ in Figure \ref{fig:gam_var}. Figure \ref{fig:gam_0} reveals a highly undesirable property of the standard PARAFAC in high dimensions, where the entire distribution of $\norm{ \bfb_{main} }_1$ shifts to the right with increasing $p$, with $\bbE \norm{ \bfb_{main} }_1 \asymp p$. The induced prior clearly lacks any automatic multiplicity adjustment property \citep{scott2010bayes}, and would bias inferences for moderate to large values of $p$. On the other hand, under the sp-PARAFAC model, the induced prior on $\norm{ \bfb_{main} }_1$ is robust to increasing $p$, as evident from Figure \ref{fig:gam_var}. The choice $\gamma = \beta p$ essentially forces a constant proportion of the variables to be assigned to the null group; see \cite{castilloneedles} for a similar choice of the hyper-parameter in a regression setting. 

\begin{table}[htbp]
  \centering
  \caption{Summary statistics of induced priors on coefficients in log-linear model parameterization.}
    \begin{tabular}{rrrrrrrr}
    \addlinespace
    \toprule
    $\gamma$ &Coefficient & Mean  & Std.dev & Min   & Max   & Skewness & Kurtosis \\
    \midrule
1   & $\beta_1$      & 0.014 & 0.831 & -6.765 & 6.389 & 0.210 & 9.109 \\
1   & $\beta_{12}$     & -0.002 & 0.340 & -2.895 & 3.105 & -0.025 & 16.583 \\
1   & $\beta_{123}$    & 0.002 & 0.196 & -2.223 & 2.632 & 0.525 & 24.686 \\
5    &$\beta_1$      & -0.002 & 0.485 & -5.648 & 5.433 & 0.031 & 27.980 \\
5    &$\beta_{12}$     & 0.000 & 0.124 & -2.085 & 2.244 & 0.495 & 93.438 \\
5   &$\beta_{123}$    & 0.000 & 0.051 & -1.214 & 0.745 & -3.701 & 159.360 \\
20    &$\beta_1$      & 0.002 & 0.246 & -3.109 & 5.669 & 2.474 & 99.554 \\
20    &$\beta_{12}$     & 0.000 & 0.042 & -1.126 & 1.819 & 9.488 & 632.790 \\
20    &$\beta_{123}$   & 0.000 & 0.009 & -0.664 & 0.214 & -44.051 & 3014.000 \\
    \bottomrule
    \end{tabular}
  \label{tab:gammtab}
\end{table}

\section{Posterior concentration}

\subsection{{\bf Preliminaries}}

In this section, we provide theoretical justification to the proposed sp-PARAFAC procedure in high dimensional settings by studying the concentration properties of the posterior with growing sample size. When the parameter space is finite dimensional, it is well known that the posterior contracts at the parametric rate of $n^{-1/2}$ under mild regularity conditions \citep{ghosal2000convergence}. However, we are interested in the asymptotic framework of the dimension $p = p_n$ growing with the sample size $n$, potentially at a faster rate, reflecting the applications we are interested in. There is a small but increasing literature on asymptotic properties of Bayesian procedures in models with growing dimensionality, with most of the focus being on linear models or generalized linear models belonging to the exponential family; refer to \cite{ghosal1999as, ghosal2000as,belitser2003adaptive,jiang2007bayesian,armagan2013posterior,bontemps2011bernstein,castilloneedles,yang2013bayesian} among others. In all these cases, the object of interest is a vector of high-dimensional regression coefficients or more generally, the conditional distribution $f(y \mid x)$ of a univariate response $y$ given high-dimensional predictors $x$. However, our object of interest is significantly different as we are concerned with estimation of the high-dimensional joint probability tensor $\pi$. 

Let $\m F_n$ denote the class of all $d_1 \times \ldots \times d_{p_n}$ probability tensors; we shall assume $d_1 = \ldots = d_{p_n} = d$ in the sequel for notational convenience. Let $\pi^{(0n)} \subset \m F_n$ be a sequence of true tensors. 
We observe $y_1, \ldots, y_n \sim \pi^{(0n)}$ and set $\bfy^{(n)} = (y_1, \ldots, y_n)$. We denote the prior distribution on $\m F_n$ induced by the sp-PARAFAC formulation by $\bbP_n$ and the corresponding posterior distribution by $\bbP_n( \cdot \mid \bfy^{(n)})$.  

For two probability tensors $\pi^{(1)}$ and $\pi^{(2)} \in \m F_n$, the $L_1$ distance is defined as:
$$
\| \pi^{(1)} - \pi^{(2)} \|_1 = \sum_{c_1 = 1}^{d} \ldots \sum_{c_{p_n}=1}^{d} | \pi_{c_1\ldots c_{p_n}}^{(1)} - \pi_{c_1 \ldots c_{p_n}}^{(2)} |. 
$$
For a sequence of numbers $\epsilon_n \to 0$ and a constant $M > 0$ independent of $\epsilon_n$, let
\begin{align}\label{eq:nbd}
U_n = \{ \pi : \| \pi - \pi^{(0n)} \|_1 \leq M \epsilon_n\}
\end{align}
denote a ball of radius $M \epsilon_n$ around $\pi^{(0n)}$ in the $L_1$ norm. We seek to find a minimum possible sequence $\epsilon_n$ such that 
\begin{align}\label{eq:post_prob}
\lim_{n \to \infty} \bbP_n( U_n^c \mid \bfy^{(n)}) \to 0, \quad \mbox{a.s.} \, \pi^{(0n)}. 
\end{align}

\subsection{\bf{Assumptions}}

In this section we state our assumptions on the true data generating model and briefly discuss their implications.  
\begin{ass} \label{ass:truth} 
The true sequence of probability tensors $\pi^{(0n)}$ are of the form 
\begin{equation}\label{eq:true_seq}
\pi^{(0n)}_{c_1 \ldots c_{p_n}} = \sum_{h=1}^{k_n} \nu_{0h} \prod_{j \in S_{0h}} \lambda_{h c_j}^{(0j)} \prod_{j \in S_{0h}^c} \lambda_{0c_j}^{(j)}, \quad 1 \leq c_j \leq d, 1 \leq j \leq p_n,  \tag{\bf{A0}}
\end{equation}
where $\lambda_0^{(j)} \in \m S^{d-1}$ are assumed to be known. Unless otherwise specified, we shall assume $\lambda_0^{(j)} = (1/d, \ldots, 1/d)$ is the probability vector corresponding to the uniform distribution on $\{1, \ldots, d\}$. 
\end{ass}
We now provide some intuition for assumption {\bf (A0)}. Letting $S_0 = \cup_{h=1}^{k_n} S_{0h}$, we can rewrite the expansion of $\pi^{(0n)}$ in {\bf (A0)} as 
\begin{align}\label{eq:true_seq_alt}
\pi^{(0n)}_{c_1 \ldots c_{p_n}} = \sum_{h=1}^{k_n} \nu_{0h} \prod_{j \in S_0} \bar{\lambda}_{h c_j}^{(0j)} \prod_{j \in S_0^c} \lambda_{0c_j}^{(j)},
\end{align}
where 
$$\bar{\lambda}_h^{(0j)} = \begin{cases} \lambda_h^{(0j)} & \text{if $j \in S_{0h}$,}
\\
\lambda_0^{(j)} &\text{if $j \in S_0 \backslash S_{0h}$.}
\end{cases}
$$
In \eqref{eq:true_seq_alt}, the term $\prod_{j \in S_0^c} \lambda_{0c_j}^{(j)}$ doesn't involve $h$ and can be factored out completely. Assumption {\bf (A0)} thus posits that the variables in $S_0^c$ are marginally independent and the entire dependence structure is driven by the variables in $S_0$. We shall refer to $S_0$ and $S_0^c$ as the non-null and null group of variables respectively. 

Let $q_n = |S_0|$ and define a mapping $j \to e_j$ from $\{1, \ldots, q_n\}$ to the ordered elements of $S_0$, so that $e_1 \leq \ldots e_{q_n}$. As $j$ varies between $1$ to $q_n$, $e_j$ ranges over the elements of $S_0$. Denote by $\psi^{(0n)}$ the $d^{q_n}$ joint probability tensor for the variables $\{ y_{ij} : j \in S_0\}$, so that
\begin{align}\label{eq:non_null}
\psi^{(0n)}_{c_1 \ldots c_{q_n}}  = \mbox{Pr}(y_{ie_1} = c_1, \ldots, y_{ie_{q_n}} = c_n) = \sum_{h=1}^{k_n} \nu_{0h} \prod_{j =1}^{q_n} \bar{\lambda}_{h c_j}^{(0e_j)}. 
\end{align}
Thus, after factoring out the marginally independent variables in $S_0^c$, {\bf (A0)} implies a standard PARAFAC expansion \eqref{eq:non_null} for $\psi^{(0n)}$ with $k_n$ many components. Since any non-negative tensor admits a standard PARAFAC distribution \citep{limcomon09}, we can always write an expansion of $\psi^{(0n)}$ as in \eqref{eq:non_null}. 

The next set of assumptions are provided below.\footnote{For sequences $a_n, b_n$, we write $a_n = o(b_n)$ if $a_n/b_n \to 0$ as $n \to \infty$ and $a_n = O(b_n)$ if $a_n \leq C b_n$ for all large $n$.}

\begin{ass}\label{ass:1}  In addition to {\bf (A0)}, $\pi^{(0n)}$ satisfies 
\item[{\bf(A1)}]
The number of components $k_n = O(1)$. 
\item[\bf{(A2)}]
Letting $s_n = \max_{1 \leq h \leq k_n} |S_{0h}|$, one has $s_n  = o(\log p_n)$. 
\item[\bf{(A3)}] 
There exists a constant $\varepsilon_0 \in (0, 1)$ such that $\lambda_{hc}^{(0j)} \geq \varepsilon_0$ for all $1 \leq h \leq k_n, 1\leq c \leq d, j \in S_{0h}$. 
\end{ass}
{\bf (A1)} and {\bf(A2)} imply that the size of the non-null group is much smaller than $p_n$, since $q_n = |S_0| \leq \sum_{h=1}^{k_n} |S_h| \leq k_n s_n \ll p_n$. 

Some discussion is in order for condition {\bf(A3)}.  First, note that we can choose $\varepsilon_0$ in a way so that $\bar{\lambda}_{hc}^{(0j)} \geq \varepsilon_0$ for all $h, c$ and $j \in S_0$. Hence, {\bf(A3)} implies a lower bound on the joint probability $\psi^{(0n)}$ in \eqref{eq:non_null}.  Such a lower bound on a compactly supported target density is a standard assumption in Bayesian non-parametric theory; see for example \cite{van2008rates}. However, unlike univariate or multivariate density estimation in fixed dimensions where the density can be assumed to be bounded below by a constant, we need to precisely characterize the decay rate of the lower bound of the joint probability. 
Since $\psi^{(0n)}$ is a $d^{q_n}$ probability tensor, $\min_{c_1, \ldots c_{s_n}} \psi_{c_1 \ldots c_{s_n}}^{(0n)} \leq (1/d)^{q_n} = \exp(- s_n k_n \log d)$. 
 Assumption {\bf (A3)} implies that 
\begin{align}\label{eq:A3hat}
\min_{c_1, \ldots c_{s_n}} \psi_{c_1 \ldots c_{s_n}}^{(0n)} \geq \exp(- q_n \log(1/\varepsilon_0)) = \exp(- c_0 s_n)
\end{align} 
for some constant $c_0 > 0$. 

\subsection{{\bf Main result}}
We are now in a position to state a theorem on posterior convergence rates. 
\begin{theorem}\label{thm:postcon}
Assume the true sequence of tensors $\pi^{(0n)} \in \m F_{n}$ satisfy assumptions {\bf (A0)} -- {\bf (A3)} and $s_n \log p_n/n \to 0$.  Also, assume the sp-PARAFAC model is fitted with the stick-breaking prior truncated to $k_n$ many components and $\gamma = \beta p_n^2$ for some constant $\beta \in (0, 1)$ in \eqref{eq:spPARAFAC}. Then, \eqref{eq:post_prob} is satisfied with $\epsilon_n = \sqrt{s_n \log p_n/n}$ in \eqref{eq:nbd}. 
\end{theorem}
A proof of Theorem \ref{thm:postcon} can be found in the appendix.  As an implication of Theorem \ref{thm:postcon}, if $p_n = n^d$ for some constant $d$, then the posterior contracts at the near parametric rate $\sqrt{(\log n)^c/n}$ for some constant $c > 0$. Moreover, consistent estimation is possible even if $p_n$ is exponentially large as long as $p_n \leq \exp(\sqrt{n})$. In particular, with $p_n = \exp(n^{\delta/2})$ for $\delta < 1$, the posterior contracts at least at the rate $n^{-(1 - \delta)/2}$. 

\begin{remark}\label{rem:factor}
We assume the number of components $k_n$ known in Theorem \ref{thm:postcon} for ease of exposition, with our main focus on dimensionality reduction. Adapting to an unknown number of components in mixture models is a well -studied problem; see, for example, \cite{ge2006consistency,pati2013jmva,shen2011adaptive}. For the infinite stick-breaking prior on the mixture components, one can use the sieving technique developed in \cite{pati2013jmva} to estimate deviation bounds for the tail sum of a stick-breaking process. 
\end{remark}

\begin{remark}\label{rem:gamma}
In practice, we recommend the choice $\gamma = \beta p_n$ for numerical stability, with $\beta = 0.2$ used as a default choice in all our examples. The probability mass function of the induced beta-bernoulli prior on $|S_h|$ with $\gamma = \beta p_n^2$ behaves like $\exp(-c s \log p_n)$ for small $s$, while the same is $\exp(-c s)$ for $\gamma = \beta p_n$; refer to the proof of Theorem \ref{thm:postcon} for further details. 
\end{remark}


\section{Posterior Computation}
Under model (\ref{eq:spPARAFAC}), we can easily proceed to draw posterior samples from a Gibbs sampler since all the full conditionals have recognizable forms. The algorithm iterates through the following steps:
\begin{enumerate}
	\item For each $j^{th}$ variable and latent class $h$, update $\lambda_h^{(j)}\equiv (\lambda_{h1}^{(j)},\ldots,\lambda_{hd_j}^{(j)})$ from a mixture of two distributions with different weights. Given the prior we specified for $\lambda_h^{(j)}$ in (\ref{eq:spPARAFAC}), the posterior maintains its conjugacy and comes from either a Dirichlet or the baseline category. i.e., for $j=1,\ldots,p$, $h=1,\ldots,k^*$, where $k^*=\mbox{max}\{z_1,\ldots,z_n\}$:
	\begin{eqnarray}
			(\lambda_h^{(j)}|-) &=& w_{0h}^{(j)} \delta_{\lambda_0^{(j)}} + 
			w_{1h}^{(j)} \mbox{Diri} \bigg(a_{j1}+\sum_{i=1}^n 1(y_{ij}=1,z_i=h),\nonumber \\
				&& \qquad \qquad \ldots,a_{jd_j}+\sum_{i=1}^n 1(y_{ij}=d_j,z_i=h)\bigg), \label{eq:postLambda}
	\end{eqnarray}		
	where $w_{0h}^{(j)}$ and $w_{1h}^{(j)}$ are the mixture weights:
	\begin{eqnarray} 
	w_{0h}^{(j)}&=& \frac{(1-\tau_h) \prod_{c=1}^{d_j} \lambda_{0c}^{(j) \sum_{i=1}^n 1(z_i=h,y_{ij}=c)}}  {(1-\tau_h) \prod_{c=1}^{d_j} \lambda_{0c}^{(j) \sum_{i=1}^n 1(z_i=h,y_{ij}=c)} + \tau_h \frac{\Gamma(\sum_{c=1}^{d_j}a_{jc})}{\prod_{c=1}^{d_j} \Gamma(a_{jc})} \cdot \frac{\prod_{c=1}^{d_j}\Gamma\big(a_{jc}+\sum_{i=1}^n 1(z_i=h,y_{ij}=c)\big)}{\Gamma\big(\sum_{c=1}^{d_j} a_{jc}+\sum_{i=1}^n 1(z_i=h)\big)}},\nonumber \\
	w_{1h}^{(j)}&=&1-w_{0h}^{(j)}.\nonumber 
	\end{eqnarray}
							
	\item 
	Let $S_{hj}$ be the allocation variable with $S_{hj}=0$ if $\lambda_h^{(j)}$ is updated from the baseline component, and $S_{hj}=1$ if $\lambda_h^{(j)}$ is from a Dirichlet posterior distribution. Update $\tau_h$, $h=1,\ldots,k^*$ from a Beta full conditional:
	\begin{eqnarray}
			\tau_h|- \sim \mbox{Beta}\bigg(1+\sum_{j=1}^p 1(S_{hj}=1), \gamma+\sum_{j=1}^p 1(S_{hj}=0)\bigg). \label{eq:postPi}
	\end{eqnarray}		
	
	\item The full conditional of $V_h$, $h=1,\ldots,k^*$ only requires the updated information on latent class allocation for all subjects:
	\begin{eqnarray}
			V_h|- \sim \mbox{Beta}\bigg(1+\sum_{i=1}^n 1(z_i=h), \alpha+\sum_{i=1}^n 1(z_i>h)\bigg). \label{eq:postV}
	\end{eqnarray}	
	
	\item We sample $z_i$, $i=1,\dots,n$ from the multinomial full conditional with:	
	\begin{eqnarray}
			\mbox{Pr}(z_i=h|-) = \frac{\nu_h\prod_{j=1}^p\lambda_{hy_{ij}}^{(j)}}{\sum_{l=1}^{k^*}\nu_l\prod_{j=1}^p\lambda_{ly_{ij}}^{(j)}}, \label{eq:postZ}
	\end{eqnarray}
	where $\nu_h=V_h\prod_{l<h}(1-V_l)$.
	
	\item Update $\alpha$ from the Gamma full conditional:
	\begin{eqnarray}
			\alpha|- \sim \mbox{Gamma}\bigg(a_{\alpha}+k^*, b_{\alpha}-\sum_{h=1}^{k^*}\mbox{log}(1-V_h)\bigg). \label{eq:postAlpha}
	\end{eqnarray}

\end{enumerate}
These steps are simple to implement and we gain efficiency by updating the parameters in blocks. For example, instead of updating $\lambda_h^{(j)}$ one at a time, we sample $\boldsymbol \lambda \equiv \{\lambda_h^{(j)}, h=1,\ldots,k^*, j=1,\ldots,p\}$ jointly with corresponding parameters in matrix form. In all our examples, we ran the chain for $25,000$ iterations, discarding the first $10,000$ iterations as burn-in and collecting every fifth sample post burn-in to thin the chain. Mixing and convergence were satisfactory based on the examination of trace plots and the run time scaled linearly with $n$ and $p$. We also carried out sensitivity analysis by multiplying and dividing the hyperparamaters $a_{\alpha}, b_{\alpha}$ and $\gamma$ in \eqref{eq:spPARAFAC} by a factor of $2$, with the conclusions remained unchanged from the default setting $a_{\alpha} = b_{\alpha} = 1$ and $\gamma = 0.2 ~ p$. 


\section{Simulation Studies}

\subsection{\bf{Estimating sparse interactions}}

\begin{figure}[htbp]
	\centering	\includegraphics[width=6.5in]{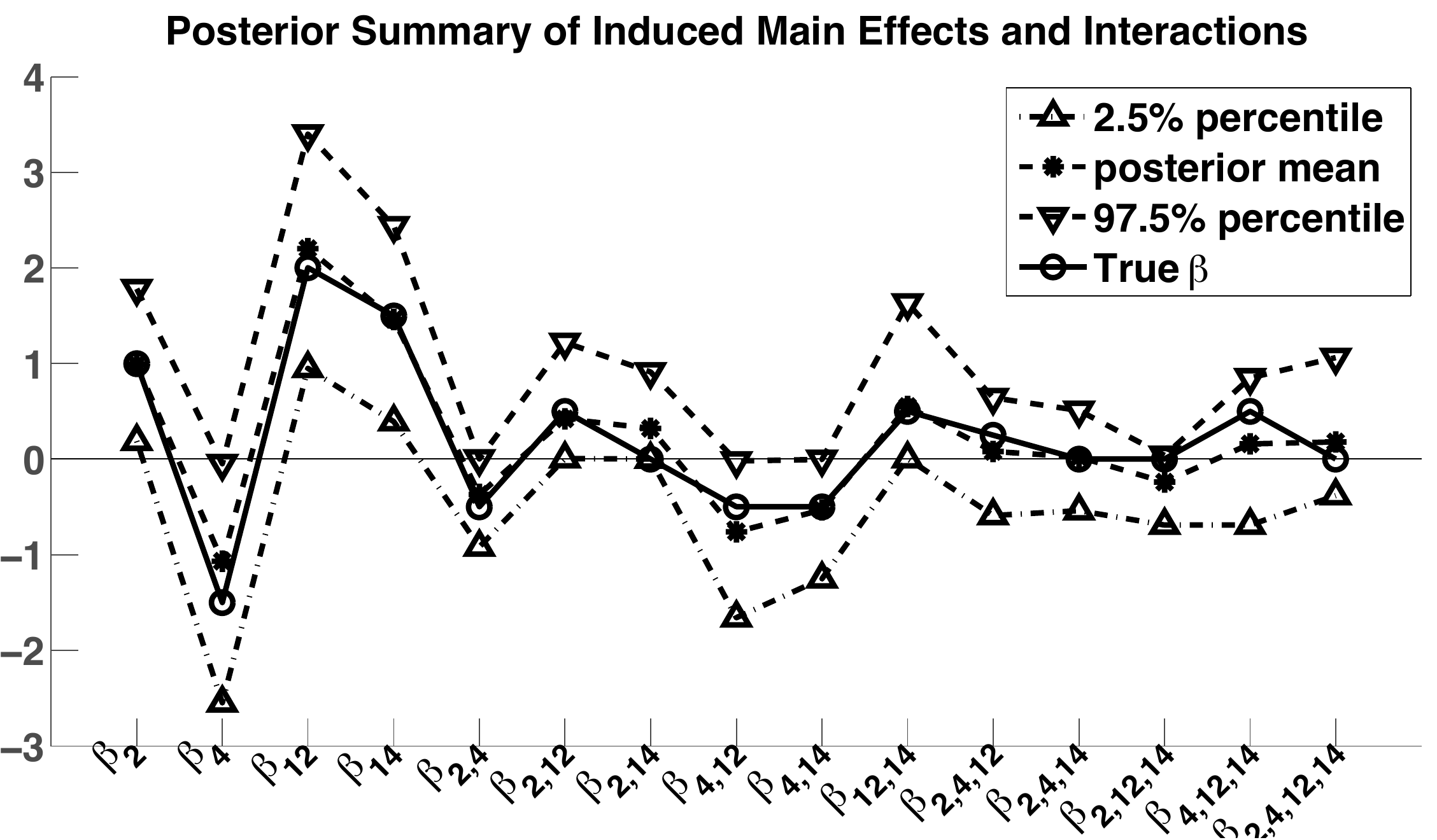}	
	\caption{Posterior means and $95\%$ credible intervals for all main effects and interactions in $S^*$ compared with the true coefficients.}
	\label{fig:beta}
\end{figure}

We first conduct a replicated simulation study to assess the estimation of sparse interactions using the proposed sp-PARAFAC model. We simulated $100$ dependent binary variables $y_{ij} \in \{0, 1\}, j = 1, \ldots, p  =100$ ($d_j = d = 2$) for $i = 1, \ldots, n = 100$ subjects from a log-linear model having up to three-way interactions:
\begin{align}\label{eq:model_lglin}
\log \bigg(\frac{\pi_{c_1\ldots c_p}}{\pi_{0\ldots 0}}\bigg) = \sum_{s=1}^{3} \sum_{S \subset \{1, \ldots, p\}: |S| = s} \beta_S 1_{(c_S = 1)}. 
\end{align}
For example, if $S = \{1, 2, 4\}$, then $\beta_S = \beta_{1,2,4}$ and $1_{(c_S = 1)} = 1_{(c_1 = 1, c_2 = 1, c_4 = 1)}$ with $1_{(\cdot)}$ denoting the indicator function. To mimic the situation where only a few interactions are present, we restrict to $S \subset S^* = \{2, 4, 12, 14\}$ and set all interactions except 
$$
\bfb = (\beta_2, \beta_4, \beta_{12}, \beta_{14}, \beta_{2,4}, \beta_{2, 12}, \beta_{4,12}, \beta_{4,14}, \beta_{12,14}, \beta_{2,4,12}, \beta_{4,12,14})^{\T}
$$
to zero. This data generating mechanism induces dependence among the variables in $S^*$, while rendering the other variables to be marginally independent. Figure~\ref{fig:beta} reports the posterior means and $95 \%$ credible intervals for all main effects and interactions for the variables in $S^*$ averaged across $100$ simulation replicates along with the true coefficients.  As illustrated in Figure~\ref{fig:beta}, averaging across the simulation replicates and different parameters, the 95\% credible intervals cover the true parameter values 80\% of the time.

\begin{figure}[htbp]
	\centering	\includegraphics[width=7in]{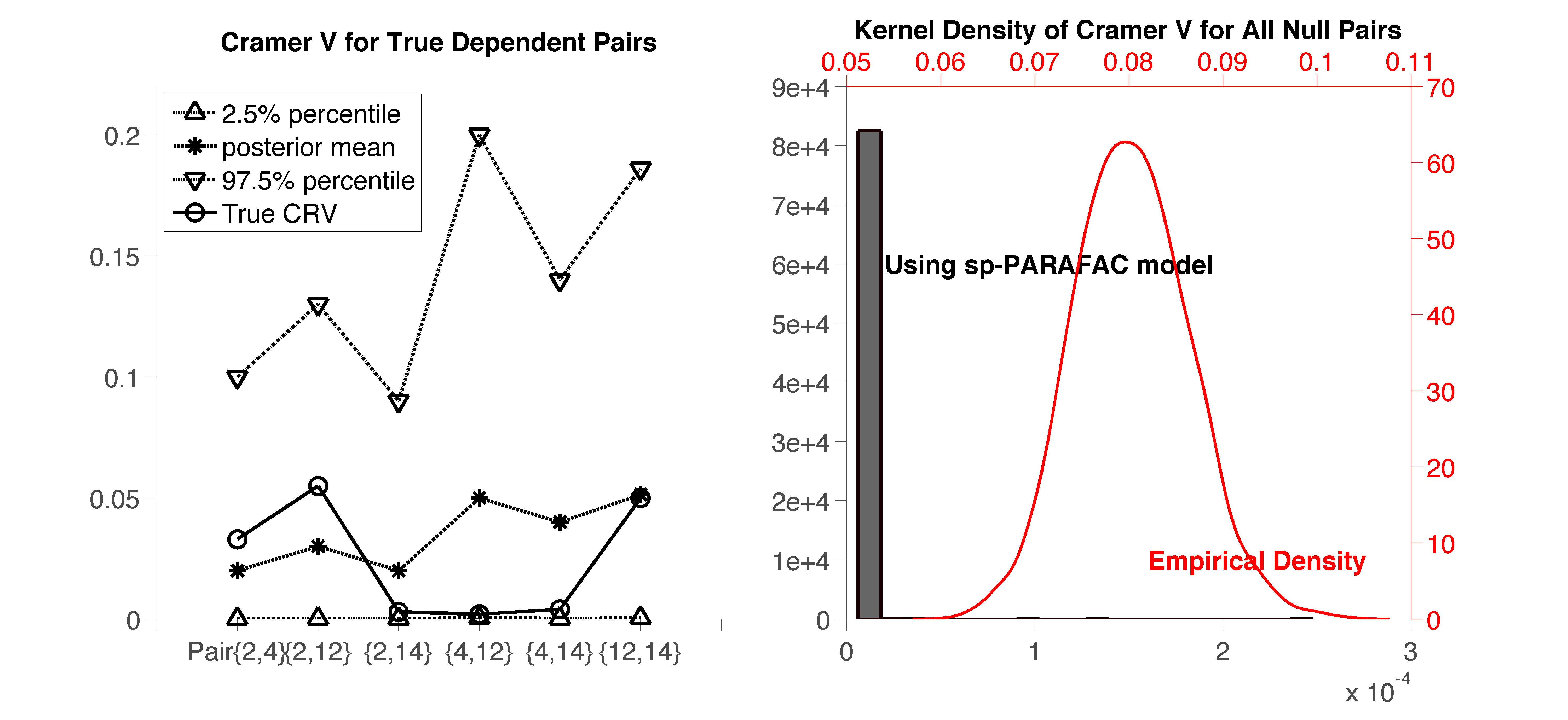}	
	\caption{Left: Posterior summaries of the Cramer's V values for all dependent pairs vs. the true Cramer's V values; Right: Estimated density of Cramer's V combining all null pairs under sp-PARAFAC vs. empirical estimation.}
	\label{fig:crv}
\end{figure}

Next, we study performance in estimating the dependence structure.  Cramer's V is a popular statistic measuring the strength of association or dependence between two (nominal) categorical variables in a contingency table, ranging from  0 (no association) to 1 (perfect association). Let $\rho_{jj'}$ denote the Cramer's V statistics for variables $j$ and $j'$, so that
	\begin{eqnarray}\label{eq:crv}
			\rho_{jj'}^2 = \frac{1}{\mbox{min}\{d_j,d_{j'}\}-1} \sum_{c_j=1}^{d_j} \sum_{c_{j'}=1}^{d_{j'}} \frac{(\pi_{c_jc_{j'}}^{(j j')} - \pi_{c_j}^{(j)} \pi_{c_{j'}}^{(j')})^2 }{ \pi_{c_j}^{(j)} \pi_{c_{(j')}}^{(j')} }, \label{eq:crv}
	\end{eqnarray}
where $\pi_{ll'}^{(j j')} = \mbox{Pr}(y_{ij} = l, y_{ij'} = l')$ and $\pi_{l}^{(j)} = \mbox{Pr}(y_{ij} = l)$. Under the log-linear model \eqref{eq:model_lglin}, $\rho = (\rho_{jj'})$ is a sparse matrix with the Cramer's V for all pairs except those in $S^* \times S^*$ being zero. This is an immediate consequence of the fact that if $(j, j') \notin S^* \times S^*$, then $y_{ij}$ and $y_{ij'}$ are independent. 

We compare estimation of the off-diagonal entries of $\rho$ under the sp-PARAFAC model with the empirical Cramer's V matrix $\hat{\rho}$. We can clearly convert posterior samples for the model parameters to posterior samples for $\rho_{j j'}$ through \eqref{eq:crv}. The empirical estimator is obtained by replacing $\pi_{c_jc_{j'}}^{(j j')}$ and $\pi_{c_j}^{(j)}$ by their empirical estimators. The left panel in Figure~\ref{fig:crv} shows the posterior summaries (averaged across simulation replicates) of the Cramer's V values for all possible dependent pairs along with the true Cramer's V values (which can be calculated from \eqref{eq:model_lglin}). In the right panel of Figure~\ref{fig:crv}, we overlay kernel density estimators of posterior samples (in grey) and the empirical estimators (in red) of the Cramer's V values for all null pairs across all simulation replicates. Note the axes are also marked in grey and red for the respective cases. The sp-PARAFAC method clearly outperforms the empirical estimator convincingly, with the posterior density for the null pairs highly concentrated near zero while the empirical estimator has a mean Cramer's V value of $0.08$ across the null pairs. 

Furthermore, we can obtain power for any non-null variable or type I error for any null variable by computing the percentage of detected significance over the simulation replicates. We first look at the power and type I error of the main effects and interactions in $S^*$, most of the power and type I error are appealing, although a few of them are far from satisfactory (see Table~\ref{tab:power} and Table~\ref{tab:typeI}). However, given the Cramer's V results in the right panel of Figure~\ref{fig:crv}, the type I error for any variable not in $S^*$ should be very small or zero. As an example, we tested the main effects and all the possible interactions for positions 20, 30, 40 and 50. The type I error rates are 0 for all of them. These results are based on examining whether 95\% intervals contain zero, and it is as expected that the approach may have difficulty assessing the exact interaction structure among a set of associated variables based on limited data.  

\begin{table}[htbp]
  \centering
  \caption{Power for Non-null Variables Based on 100 Simulations}
    \begin{tabular}{rrrrrrrrrrrr}
    \addlinespace
    \toprule
          & {\bf $\beta_{2}$} & {\bf $\beta_{4}$} & {\bf $\beta_{12}$} & {\bf $\beta_{14}$} & {\bf $\beta_{2,4}$} & {\bf $\beta_{2,12}$} & {\bf $\beta_{4,12}$} & {\bf $\beta_{4,14}$} & {\bf $\beta_{12,14}$} & {\bf $\beta_{2,4,12}$} & {\bf $\beta_{4,12,14}$} \\
    \midrule
    {\bf Power} & 0.97  & 0.9   & 1     & 1     & 0.95  & 0.99  & 0.98  & 0.97  & 0.99  & 0     & 0 \\
    {\bf True coefficient} & 1     & -1.5  & 2     & 1.5   & -0.5  & 0.5   & -0.5  & -0.5  & 0.5   & 0.25  & 0.5 \\
    \bottomrule
    \end{tabular}
  \label{tab:power}
\end{table}

\begin{table}[htbp]
  \centering
  \caption{Type I Error for Null Variables Based on 100 Simulations}
    \begin{tabular}{rrrrr}
    \addlinespace
    \toprule
          & {\bf $\beta_{2,14}$} & {\bf $\beta_{2,4,14}$} & {\bf $\beta_{2,12,14}$} & {\bf $\beta_{2,4,12,14}$} \\
    \midrule
    {\bf Type I error} & 0.97  & 0     & 0.68  & 0 \\
    {\bf True coefficient} & 0     & 0     & 0     & 0 \\
    \bottomrule
    \end{tabular}
  \label{tab:typeI}
\end{table}

\subsection{\bf{Comparison with standard PARAFAC}}

We now conduct a simulation study  to compare estimation of the Cramer's V matrix $\rho$ under the proposed approach to the usual specification of the PARAFAC model without any sparsity as in \eqref{eq:dx}, which is equivalent to setting $\gamma = 0$ in \eqref{eq:spPARAFAC}. We considered $100$ simulation replicates, with data in each replicate consisting of $p = 100$ categorical variables for $n = 100$ subjects, with each variable having $4$ possible levels ($d_j = d = 4$). Two simulation settings were considered to induce dependence between the variables in $S^* = \{2, 4, 12, 14\}$: (i) via multiple subpopulations as in the simulation study in \cite{dunsonxing08}, and (ii) via a nominal GLM model $Pr(y_{ij}=c)=\frac{exp(\boldsymbol y_{i(j)} \boldsymbol \beta_{c})}{1+\sum_{c=2}^4 exp(\boldsymbol y_{i(j)} \boldsymbol \beta_{c})}$ for $j \in S^*$, where $\boldsymbol y_{i(j)} \boldsymbol \beta_{c}$ is a linear combination of all variables that are associated with the $j^{th}$ variable excluding the $j^{th}$ variable. The remaining variables were independently generated from a discrete uniform distribution.

\begin{figure}[htbp]
	\centering
		\includegraphics[width=3in]{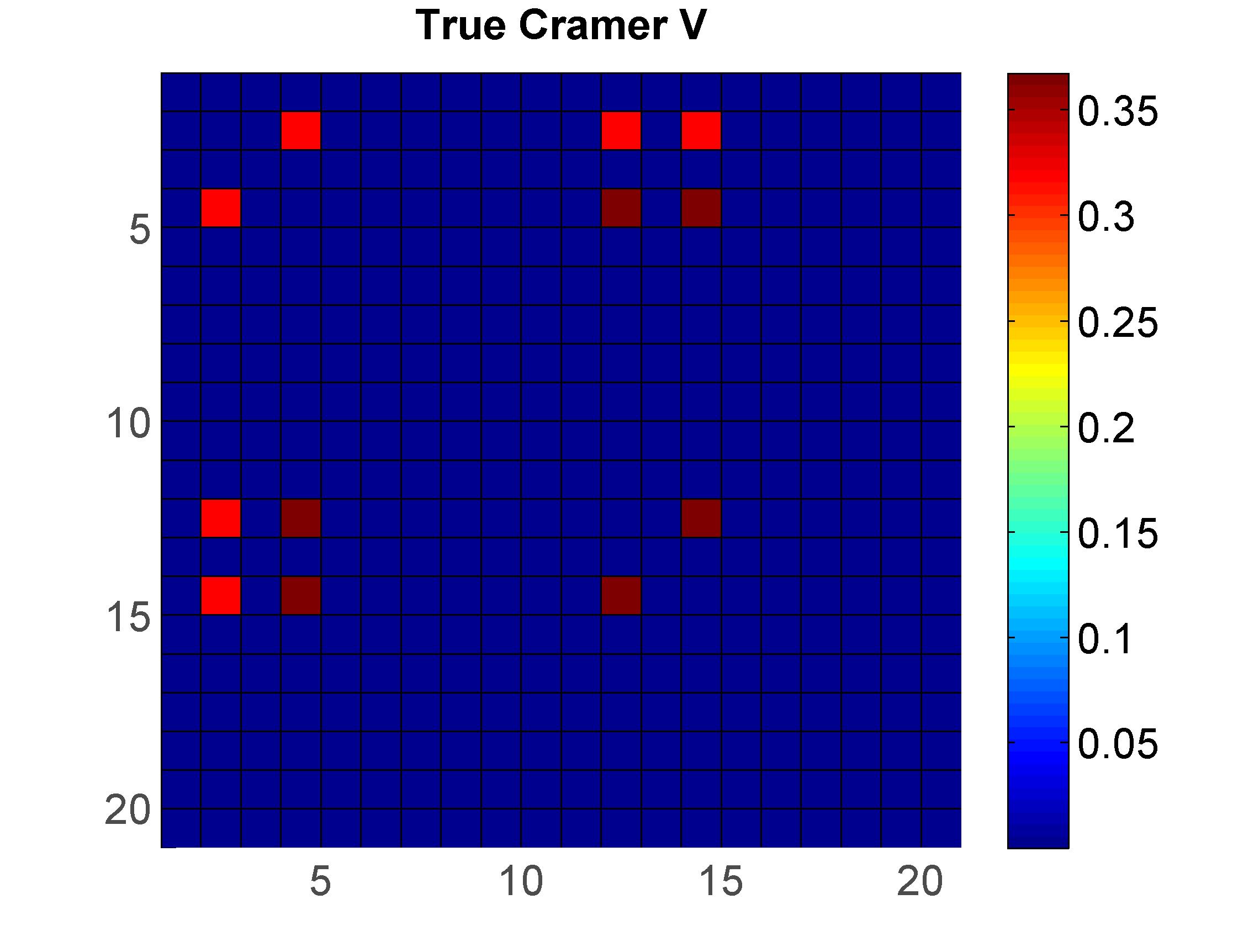}
		\includegraphics[width=3in]{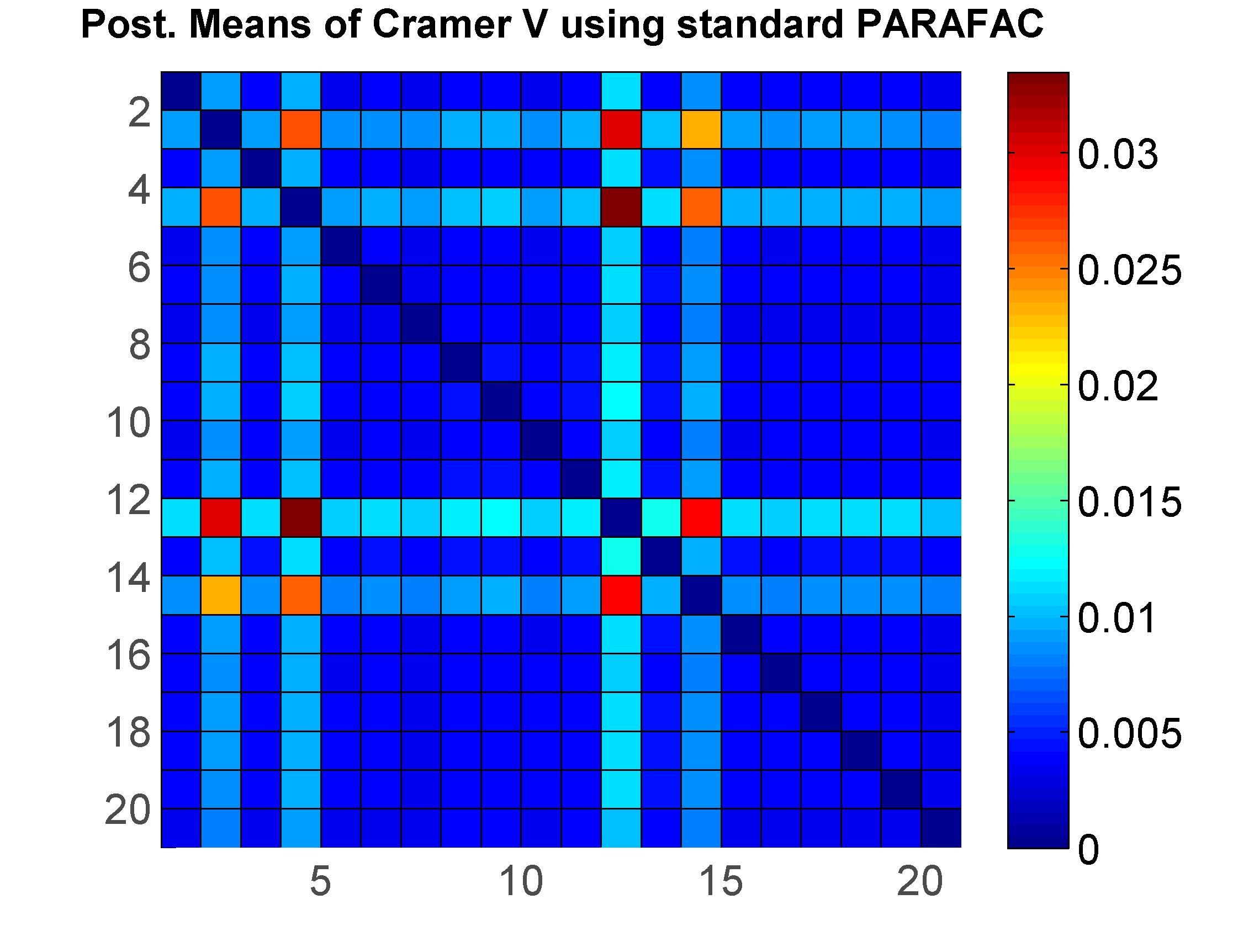}
	\caption{Simulation setting (i) -- Left: True Cramer's V matrix; Right: Posterior means of Cramer's V using standard PARAFAC.}
	\label{fig:simCRVtrue}
\end{figure}

\begin{figure}[htbp]
	\centering
		\includegraphics[width=3in]{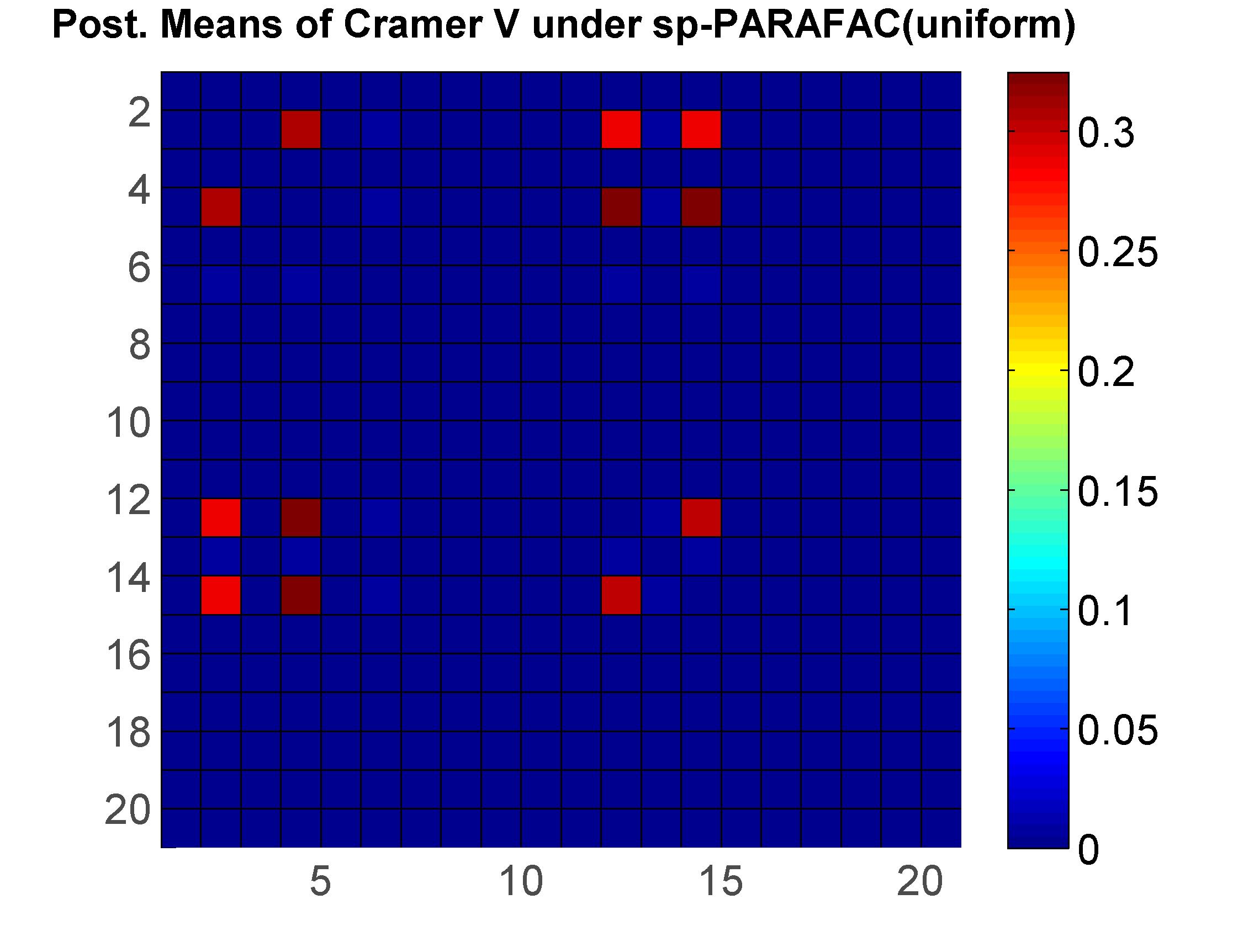}
		\includegraphics[width=3in]{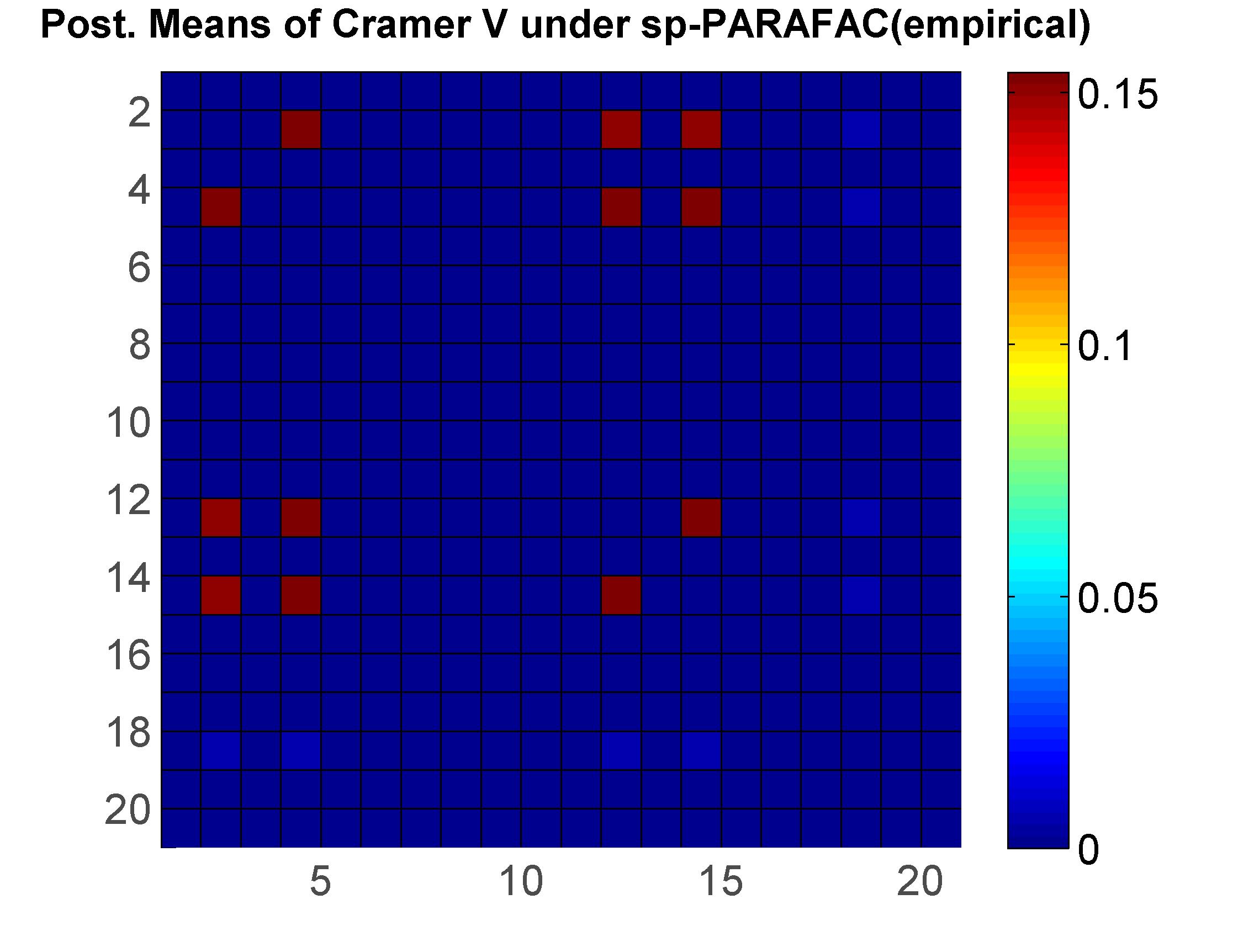}
	\caption{Posterior means of Cramer's V under simulation setting (i) using proposed method -- Left: with $\lambda_0^{(j)}$ being  discrete uniform;  Right: with $\lambda_0^{(j)}$ being empirical estimates of the marginal category probabilities.}
	\label{fig:simCRV}
\end{figure}

\begin{figure}[htbp]
	\centering
		\includegraphics[width=6.5in]{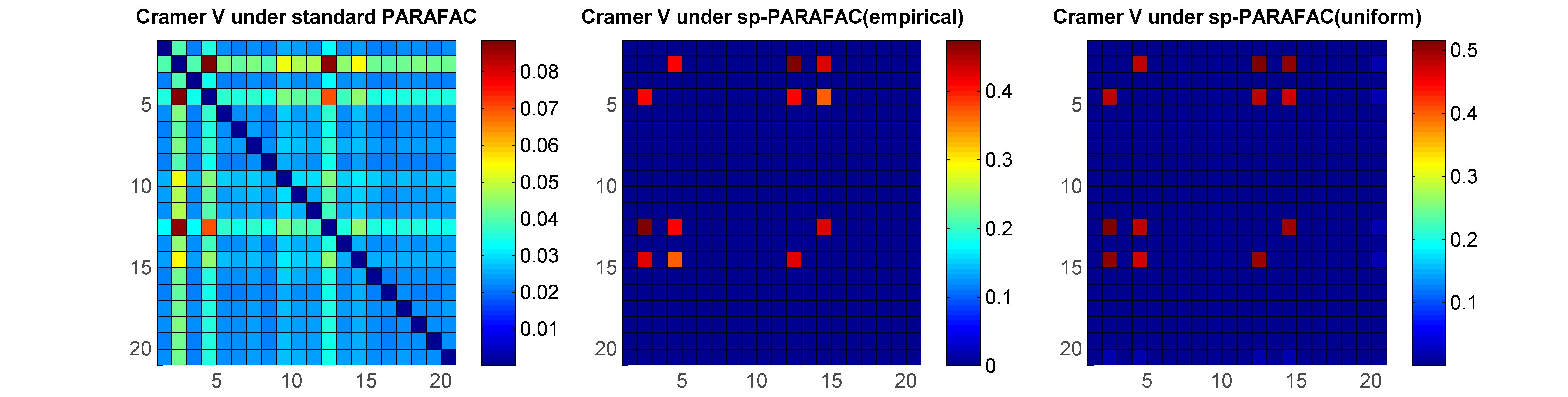}
	\caption{Posterior means of Cramer's V under simulation setting (ii) -- Left: using standard PARAFAC;  Middle: under proposed method using empirical marginal with Diri(1,...,1) prior for $\lambda_0$;   Right: using proposed method with discrete uniform $\lambda_0$.}
	\label{fig:simCRVglm}
\end{figure}

The color plot on the left in Figure~\ref{fig:simCRVtrue} shows the true pairwise Cramer's V values under simulation setting (i) (only the top-left $20 \times 20$ sub matrix of $\rho$ is shown for clarity). 
Figure~\ref{fig:simCRVtrue} (right) and Figure~\ref{fig:simCRV} represent one of the replicates, in which the right plot in Figure~\ref{fig:simCRVtrue} shows the Cramer's V under the standard non-sparse PARAFAC method, while Figure~\ref{fig:simCRV} shows the Cramer's V using our method with two different baseline components. It is obvious that our approach has much better estimates for not only the true dependent pairs but also the true nulls.   Results for simulation (ii) shown in Figure~\ref{fig:simCRVglm} again show superiority of our sparse improvement to PARAFAC.

\section{Application}
\subsection{\bf{Splice-junction Gene Sequences}}
\begin{figure}[htbp]
	\centering	\includegraphics[width=6.5in]{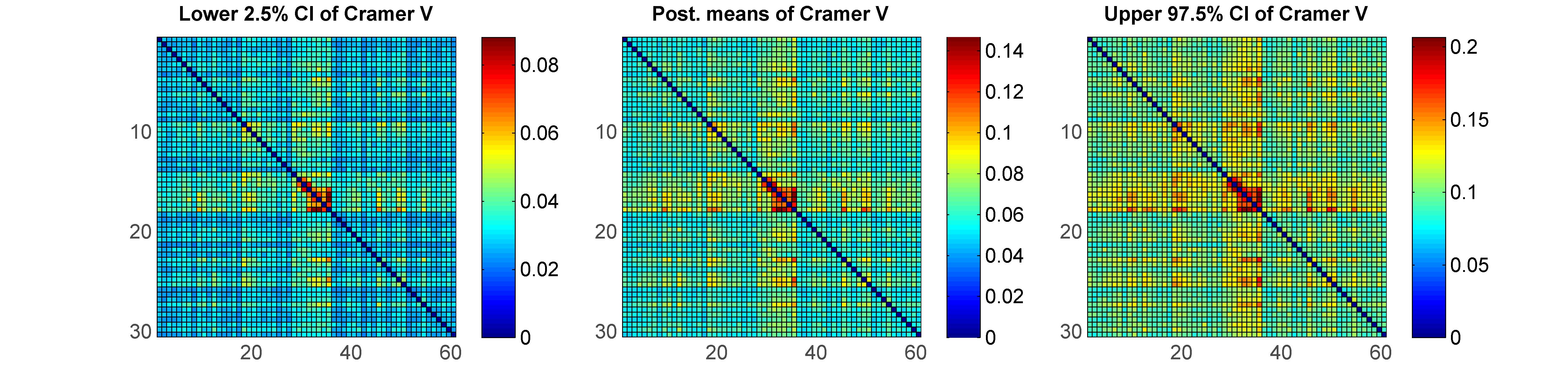}	\includegraphics[width=6.5in]{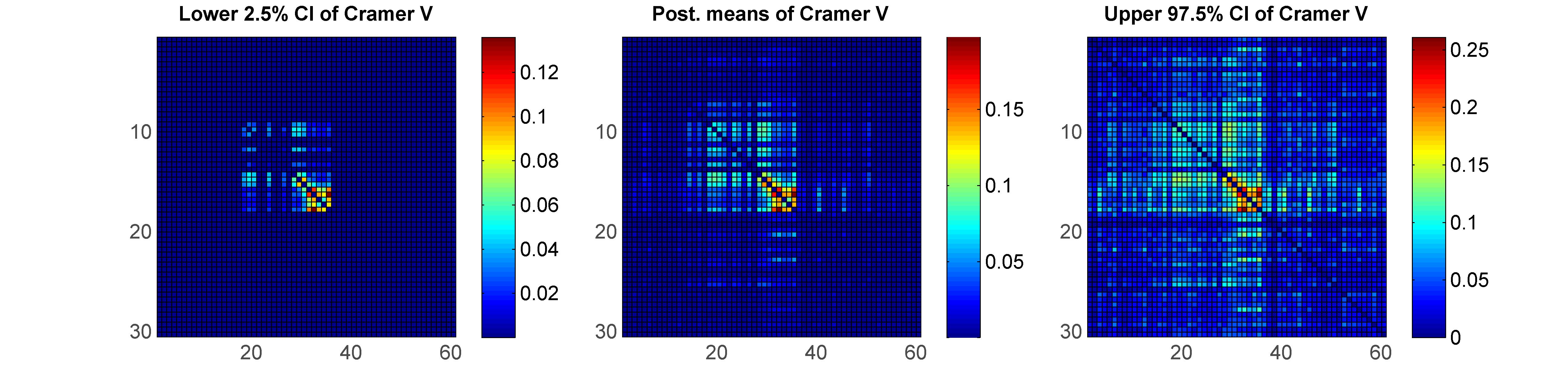}
	\caption{Posterior quantiles of Cramer's V with 120 sequences of splice data -- Upper panel: under standard PARAFAC; Bottom panel:under proposed method.}
	\label{fig:spliceCRV}
\end{figure}

\begin{figure}[htbp]
	\centering	\includegraphics[width=6.5in]{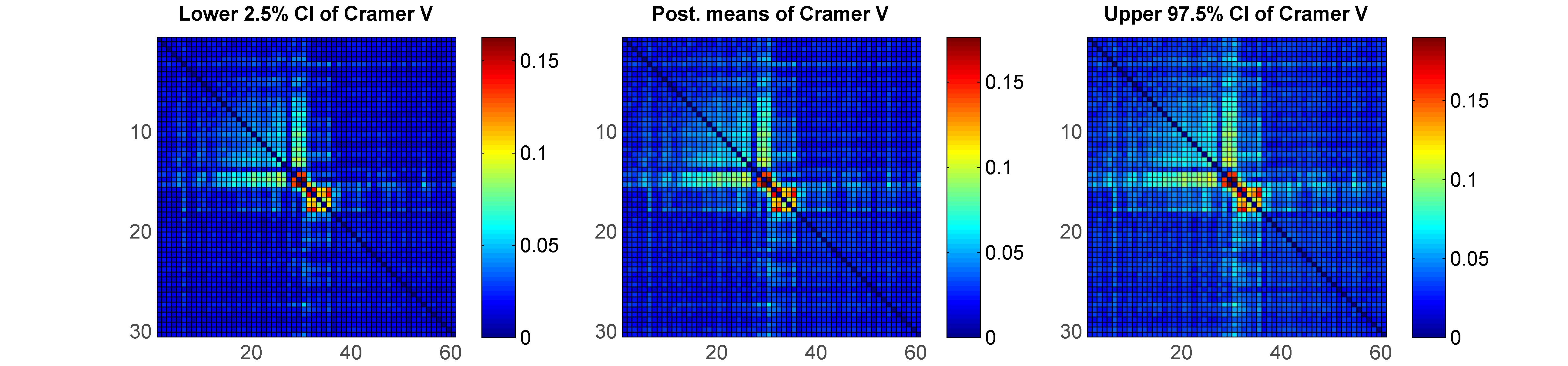}	\includegraphics[width=6.5in]{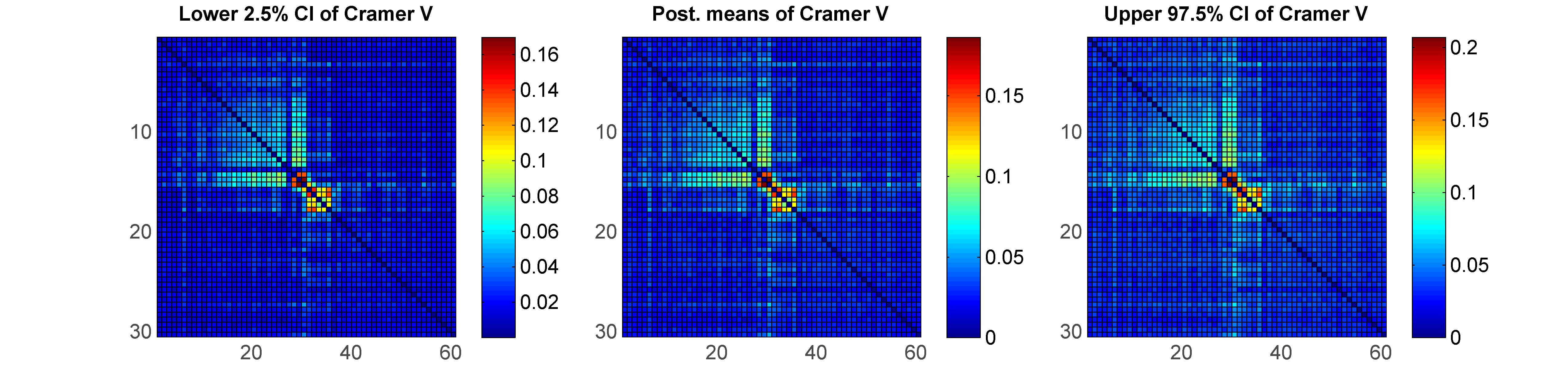}
	\caption{Posterior quantiles of Cramer's V with 3,175 sequences of splice data -- Upper panel: under standard PARAFAC; Bottom panel:under proposed method.}
	\label{fig:spliceALLCRV}
\end{figure}

We applied the method to the Splice-junction Gene Sequences (abbreviated as splice data below). Splice junctions are points on a DNA sequence at which `superfluous' DNA is removed during the process of protein creation in higher organisms. These data consist of A, C, G, T nucleotides at $p = 60$ positions for $N = 3,175$ sequences. Since its sample size is much larger than the number of variables, we compared our approach with the standard PARAFAC in two scenarios, first a small randomly selected subset (of size $n = 2p = 120$) of the full data set, and second, the full data set itself. Using two different sample sizes in this manner allows for a study of the new and existing method and a comparison to a gold standard (a sufficiently large data set). We ran the analysis to estimate the pairwise positional dependence structure under the standard PARAFAC method and the proposed approach with discrete uniform baseline component. As is apparent in Figure~\ref{fig:spliceALLCRV}, both methods have similar performance when $n\gg p$, however, in the smaller sample size situation, Figure~\ref{fig:spliceCRV} demonstrates that our proposed method has the advantage of identifying the dependence structure and pushing the independent pairs to zero, which it is closer to the results in a large sample case (Figure~\ref{fig:spliceALLCRV}).

\section{Discussion}

We have proposed a sparse modification to the widely-used PARAFAC tensor factorization, and have applied this in a Bayesian context to improve analyses of ultra sparse huge contingency tables.  Given the compelling success in this application area, we hope that the proposed notion of sparsity will have a major impact in other areas, including tensor completion problems in machine learning.  There is an enormous literature on low rank and sparse matrix factorizations, and the sp-PARAFAC should facilitate scaling of such approaches to many-way tables while dealing with the inevitable curse of dimensionality.  Although we take a Bayesian approach, we suspect that frequentist penalized optimization methods can also exploit our same concept of sparsity in learning a compressed characterization of a huge array based on limited data.

\section*{Appendix}
\subsection{\bf{Proof of Theorem \ref{thm:postcon}}}

We verify the conditions of Theorem 4 in Yang \& Dunson (2013), which is a minor modification of Theorem 2 appearing in \cite{ghosal2000convergence}. Let $\epsilon_n \to 0$ be such that $n \epsilon_n^2 \to \infty$ and $\sum_{n \geq } \exp(-n \epsilon_n^2) \leq \infty$. Suppose there exist a sequence of sets $\m P_n \subset \m F_n$ and a constant $C> 0$ such that the following hold: \footnote{Given a metric space $(\m X, d)$, let  $N(\epsilon; \m X, d)$ denote its $\epsilon$-covering number, i.e., the minimum number of $d$-balls of radius $\epsilon$ needed to cover $\m X$.}
\begin{enumerate}
\item $\log N(\epsilon_n; \m P_n, \| \cdot \|_1) \leq n \epsilon_n^2$;
\item $\bbP_n(\m F_n \cap \m P_n^c) \leq \exp\{ - (2 + C) n \epsilon_n^2 \}$; 

\item $\bbP_n \l( \pi : \l \| \log \frac{\pi}{\pi^{(0n)}} \r \|_{\infty} \leq \epsilon_n^2 \r) \geq \exp(- C n \epsilon_n^2)$.  
\end{enumerate}
Then, the posterior contracts at the rate $\epsilon_n$, i.e., \eqref{eq:post_prob} is satisfied. We now proceed to verify conditions (1) -- (3). We define, 
\begin{align}\label{eq:sieve}
\m P_n = \l\{ \pi \in \m F_n : \pi_{c_1 \ldots c_p} = \sum_{h=1}^{k_n} \nu_h^* \prod_{j \in S_h^*} \lambda_{h c_j}^{(*j)} \prod_{j \in S_h^{*c}} \lambda_{0 c_j}^{(j)}; \, \nu \in \m S^{(k_n-1)}, \, | S_h^*| \leq A s_n, h =1, \ldots, k_n \r\}
\end{align}
where $\m S^{(r-1)}$ denotes the $(r-1)$-dimensional probability simplex and $A > 0$ is an absolute constant. We shall use $C$ to denote an absolute constant whose meaning may change from one line to the next. 

To estimate $N(\epsilon_n; \m P_n, \| \cdot \|_1)$, we make use of the following Lemma, which follows in a straightforward manner by repeated uses of the triangle inequality. 

\begin{lemma}\label{lem:l1}
Let $\pi^{(1)}, \pi^{(2)} \in \m F_n$ with 
$$
\pi^{(i)} = \sum_{h=1}^{k_n} \nu_{ih} \lambda_{i h}^{(1)} \otimes \ldots \otimes \lambda_{i h}^{(p_n)}, ~ i = 1, 2. 
$$
Then,
$$
\| \pi^{(1)} - \pi^{(2)} \|_1 \leq \sum_{h=1}^{k_n} | \nu_{1h} - \nu_{2h} | + \sum_{h=1}^{k_n} \nu_{2h} \bigg(\sum_{j=1}^{p_n} \sum_{c=1}^d | \lambda_{1hc}^{(j)} - \lambda_{2hc}^{(j)} | \bigg).
$$
\end{lemma}
Lemma \ref{lem:l1} implies that if $\pi^{(1)}, \pi^{(2)} \in \m P_n$ with $S_{1h}^* = S_{2h}^* = S_h^*$, then 
$$
\| \pi^{(1)} - \pi^{(2)} \|_1 \leq \sum_{h=1}^{k_n} | \nu_{1h} - \nu_{2h} | + \sum_{h=1}^{k_n} \nu_{2h} \bigg(\sum_{j \in S_h^*}\sum_{c=1}^d | \lambda_{1hc}^{(j)} - \lambda_{2hc}^{(j)} | \bigg). 
$$
Based on the above observation, we create an $\epsilon$-net of $\m P_n$ as follows: In \eqref{eq:sieve}, (i) vary $S_h^*$ over all possible subsets of $\{1, \ldots, p_n\}$ with $|S_h^*| \leq A s_n$ for $h = 1, \ldots, k_n$, (ii) for $h \in \{1, \ldots k_n\}$ and $j \in S_h^*$, vary $\lambda_h^{(*j)}$ over an $\epsilon_n/(2A d s_n)$-net of $\m S^{(d-1)}$ and (iii) vary $\nu^*$ over an $\epsilon_n/(2 k_n)$-net of $\m S^{(k_n-1)}$. 

For a fixed $h$, there are $\sum_{s=0}^{A s_n} {p \choose s}$ subsets of size smaller then or equal to $A s_n$. Using the inequality ${p \choose s} \leq (pe/s)^s$ for $s \leq p/2$, the number of possible subsets in (i) can be bounded above by $\exp(C k_n s_n \log p_n)$. Hence,
$$
N(\epsilon_n; \m P_n, \| \cdot \|_1) \leq \exp(C k_n s_n \log p_n) \, N(\epsilon_n/(2A d s_n); \m S^{d-1}, \| \cdot \|_1)^{2 A d s_n k_n } \, N(\epsilon_n/(2 k_n); \m S^{k_n - 1}, \| \cdot \|_1). 
$$
Using the fact that $N(\delta, \m S^{r-1}, \| \cdot \|_1) \leq (C/\delta)^{r}$ \citep{vershynin2010introduction}, the right hand side in the above display can be bounded above by $\exp(C s_n \log p_n) = \exp(n \epsilon_n^2)$, since $k_n = O(1)$. 

We now bound $\bbP_n(\m F_n \cap \m P_n^c)$. Recall that in the sp-PARAFAC model, the induced prior on the subset size $| S_h|$ is $\mbox{Bin}(p_n, \tau_h)$, with $\tau_h \sim \mbox{Beta}(1, \gamma)$. Now,
$$
\bbP_n((\m F_n \cap \m P_n^c) \leq \mbox{Pr}(\exists \, h \in \{1, \ldots, k_n\} \, \mbox{s.t.} \, |S_h| \geq A s_n) \leq k_n P(|S_1| > A s_n). 
$$
Integrating $\tau_1$, the distribution of $|S_1|$ is a beta-bernoulli distribution with probability mass function
\begin{align*}
\mbox{Pr}(|S_1| = s) & = {p \choose s} \frac{1}{\mbox{B}(1, \gamma)} \int_{\tau = 0}^{1} \tau^s (1 - \tau)^{p_n-s} (1 - \tau)^{\gamma-1} d \tau \\
&=  {p_n \choose s} \frac{\mbx B(1+s, \gamma + p_n - s)}{\mbox{B}(1, \gamma)} \\
& = \frac{1}{\gamma} \frac{p_n !}{(p_n-s) !} \frac{(\gamma + p_n -s -1)!}{(\gamma + p_n)!},
\end{align*}
for $s = 0, 1, \ldots, p_n$. $\mbx B(\cdot, \cdot)$ denotes the Beta function in the above display. Hence, for $s \geq 1$, 
$$
\frac{\mbox{Pr}(|S_1| = s)}{\mbox{Pr}(|S_1| = s-1)} = \frac{(p_n-s+1)}{(p_n-s + \gamma)}. 
$$
Now, letting $\gamma = p_n^2$, one has for any $p_n \geq 2$ and $1 \leq s \leq p_n/2$, 
$$
\frac{1}{4p_n} \leq \frac{(p_n-s+1)}{(p_n-s + \gamma)} \leq \frac{1}{p_n}. 
$$
In general, for $\gamma = \beta p_n^2$, we can bound this from both sides by $C/ p_n$. 
Noting that $\mbox{Pr}(|S_1| = 0) = C/p_n^3$, we have
$$
\mbox{Pr}(|S_1| = s) = \frac{C}{p_n^3} \prod_{j=1}^s \frac{\mbox{Pr}(|S_1| = j)}{\mbox{Pr}(|S_1| = j-1)}, 
$$
implying there exists constants $c_1, c_2 > 0$ such that
\begin{align}\label{eq:bet_bern}
e^{- c_1 (s+3) \log p_n} \leq \mbox{Pr}(|S_1| = s) \leq e^{- c_2 (s+3) \log p_n},
\end{align}
for $0 \leq s \leq p_n/2$. In particular, the upper bound holds for all $0 \leq s \leq p_n$, since $(p_n - s + 1)/(p_n - s + \gamma) \leq C/p_n$ for all $s$. Hence, for $n$ large enough so that $s_n \geq 3$, 
$$
\mbx P(|S_1| > A s_n) \leq \sum_{j = A s_n  + 1}^{p_n} \exp( -C  j \log p_n) \leq \exp(- C s_n \log p_n) \leq \exp(-n \epsilon_n^2). 
$$

We finally show that (3) holds. Recall the decomposition of $\pi^{(0n)}$ from \eqref{eq:true_seq_alt}. A probability tensor $\pi$ following a sp-PARAFAC model with a truncated stick-breaking prior on $\nu$ can be parameterized as
$$
\theta_{\pi} = \l(\nu, \{S_h\}_{1\leq h \leq k_n}, \{ \lambda_h^{(j)} \}_{1\leq h \leq k_n, j \in S_h} \r),
$$
where $\nu \in \m S^{k_n-1}, S_h \subset \{1, \ldots, p_n\}, \lambda_h^{(j)} \in \m S^{d-1}$. 
Consider the following subset $\m A$ of the parameter space,
$$
\m A = \l\{ S_h = S_0, 1\leq h \leq k_n; \, \sum_{h=1}^{k_n} | \nu_h - \nu_{0h} | \leq \frac{\epsilon_n^2}{2 e^{c_0 s_n}}; \, \sum_{c=1}^d |\lambda_{hc}^{(j)} - \bar{\lambda}_{hc}^{(0j)} | \leq \frac{\epsilon_n^2 \varepsilon_0}{4 q_n}, 1\leq h \leq k_n, j \in S_0 \r\}.
$$ 
We now show that $\theta_{\pi} \in \m A$ implies $\log \| \pi/\pi^{(0n)} \|_{\infty} \leq \epsilon_n^2$, so that $\bbP_n(\log \| \pi/\pi^{(0n)} \|_{\infty} \leq \epsilon_n^2)$ can be bounded below by $\bbP_n(\m A)$. 
First, observe that since $S_h = S_0$ for all $h$ on $\m A$, $\pi/\pi^{(0n)} = \psi/\psi^{(0n)}$, where $\psi^{(0n)}$ is as in \eqref{eq:non_null} and $\psi$ is the $d^{q_n}$ joint probability tensor implied by the sp-PARAFAC model for the variables $\{ y_{ij} : j \in S_0\}$,
$$
\psi_{c_1 \ldots c_{q_n}} = \sum_{h=1}^{k_n} \nu_{h} \prod_{j \in S_0} \lambda_{h c_j}^{(e_j)}. 
$$
Hence,  
$$
\log \l \| \frac{\pi}{\pi^{(0n)}} \r\|_{\infty} =  \log \l \| \frac{\psi}{\psi^{(0n)}} \r\|_{\infty} \leq \log \l(1 + \l \| \bigg( \frac{\psi}{\psi^{(0n)}} - 1\bigg)  \r\|_{\infty} \r) \leq \l \| \bigg( \frac{\psi}{\psi^{(0n)}} - 1\bigg)  \r\|_{\infty},
$$
where the penultimate step follows from an application of triangle inequality and the last step uses $\log(1 + x) \leq x$ for $x \geq 0$. For any $c_1, \ldots, c_{s_n}$, by an application of triangle inequality,  
\begin{align}\label{eq:linfty}
| \psi_{c_1 \ldots c_{s_n}} - \psi^{(0n)}_{c_1 \ldots c_{s_n}} | \leq \sum_{h=1}^{k_n} |\nu_h - \nu_{0h}| + \sum_{h=1}^{k_n}
 \nu_{0h} ~ \big| \prod_{j=1}^{q_n} \lambda_{hc_j}^{(e_j)} - \prod_{j=1}^{q_n} \bar{\lambda}_{hc_j}^{(0e_j)} \big|. 
\end{align}

We now state a Lemma to facilitate bounding the second term of the above display. 
\begin{lemma}\label{lem:ineq}
Let $v_1, \ldots v_r \in (\varepsilon_0, 1 - \varepsilon_0)$ for some $\varepsilon_0 > 0$. Let $\delta > 0$ be such that $r \delta  < \varepsilon_0/2$. Then, if $u_1, \ldots, u_r$ satisfy $|u_j - v_j| \leq \delta$ for all $j = 1, \ldots, r$, then
$$
|u_1 \ldots u_r  - v_1 \ldots v_r | \leq \frac{2 r \delta}{\varepsilon_0}  \, v_1 \ldots v_r. 
$$
\end{lemma}
Apply Lemma \ref{lem:ineq} with $r = q_n, u_j = \bar{\lambda}_{hc_j}^{(0e_j)}$ and $\delta = \epsilon_n^2 \varepsilon_0/(4 q_n)$ (clearly $r \delta/\varepsilon_0 = \epsilon_n^2/4 < 1/2$) to obtain that for any $1\leq h \leq q_n$, 
$\big| \prod_{j=1}^{q_n} \lambda_{hc_j}^{(e_j)} - \prod_{j=1}^{q_n} \bar{\lambda}_{hc_j}^{(0e_j)} \big| \leq (\epsilon_n^2/2) \prod_{j=1}^{q_n} \bar{\lambda}_{hc_j}^{(0e_j)}$. Substituting this bound in \eqref{eq:linfty}, we have on $\m A$,
\begin{align*}
\frac{| \psi_{c_1 \ldots c_{s_n}} - \psi^{(0n)}_{c_1 \ldots c_{s_n}} |}{\psi^{(0n)}_{c_1 \ldots c_{s_n}}}  \leq \frac{\sum_{h=1}^{k_n} |\nu_h - \nu_{0h}|}{e^{-c_0 s_n}} + (\epsilon_n^2/2) \frac{\sum_{h=1}^{k_n}
 \nu_{0h} \prod_{j=1}^{q_n} \bar{\lambda}_{hc_j}^{(0e_j)} }{ \psi_{c_1 \ldots c_{s_n}}^{(0n)}} \leq \epsilon_n^2.
\end{align*}
For the two terms in the above display after the first inequality, we used the lower bound \eqref{eq:A3hat} for the first term along with $\sum_{h=1}^{k_n} | \nu_h - \nu_{0h} | \leq \epsilon_n^2/(2 e^{c_0 s_n)}$ on $\m A$, and by definition of $\psi^{(0n)}$, the second term is $\epsilon_n^2/2$. 

It thus remains to lower bound $\bbP_n(\m A)$. By independence across $h$, $\mbox{Pr} (S_h = S_0, 1\leq h \leq k_n) = \mbox{Pr}(S_1 = S_0)^{k_n}$. Further, by exchangeability of the prior on $S_1$, since all subsets of a particular size receive the same prior probability, $\mbox{Pr}(S_1 = S_0) = \mbox{Pr}(|S_1| = q_n)/{p_n \choose q_n}$. From \eqref{eq:bet_bern}, $\mbox{Pr}(|S_1| = q_n \geq \exp(- C s_n \log p_n)$. Using ${p_n \choose q_n} \leq (p_n e/q_n)^{q_n}$, we conclude that $\mbox{Pr}(S_1 = S_0) \geq \exp(-C s_n \log p_n)$. 

Recall that $\nu_h = \nu_h^* \prod_{l < h} (1 - \nu_l^*)$, where $\nu_l^* \sim \mbox{Beta}(1, \alpha)$ independently. Find numbers $\{ \nu_{0h}^*\}$ such that $\nu_{0h} = \nu_{0h}^* \prod_{l < h} (1 - \nu_{0l}^*)$. It is easy to see that there exists a constant $C>0$ such that $| \nu_h^* - \nu_{0h}^*| \leq \epsilon_n/(C k_n)$ for all $h = 1, \ldots, k_n$ implies $\sum_{h=1}^{k_n} |\nu_h - \nu_{0h}| \leq \epsilon_n$. 
Hence, using a general result on small ball probability estimate of Dirichlet random vectors (Lemma 6.1 of \cite{ghosal2000convergence}), one has
$$
\mbox{Pr}\bigg(\sum_{h=1}^{k_n} | \nu_h - \nu_{0h} | \leq \frac{\epsilon_n^2}{2 e^{c_0 s_n}}\bigg) \geq \exp\{- C s_n \log(1/\epsilon_n) \}. 
$$
Again, applying Lemma 6.1 of \cite{ghosal2000convergence},
$$
\mbox{Pr}\bigg( \sum_{c=1}^d |\lambda_{hc}^{(j)} - \bar{\lambda}_{hc}^{(0j)} | \leq \frac{\epsilon_n^2 \varepsilon_0}{4 q_n}\bigg) \geq \exp\{- C \log(s_n/\epsilon_n)\}.
$$
Combining, we get $\mbox{Pr}(\m A) \geq \exp(-C s_n \log p_n) \geq \exp(- n \epsilon_n^2)$. Hence, we have established (1) -- (3), completing the proof.

\subsection{\bf{Proof of Lemma \ref{lem:ineq}}}

Observe that 
$$
| u_1 \ldots u_r - v_1 \ldots v_r | = |v_1 \ldots v_r| \l | \frac{u_1 \ldots u_r}{v_1 \ldots v_r} - 1 \r| = v_1 \ldots v_r \max \l \{\frac{u_1 \ldots u_r}{v_1 \ldots v_r} - 1, 1 - \frac{u_1 \ldots u_r}{v_1 \ldots v_r} \r\}.
$$
Now, since $u_h \leq v_h + \delta$ for all $h$, 
$$
\frac{u_1 \ldots u_r}{v_1 \ldots v_r} \leq \prod_{h=1}^r (1 + \delta/v_h) \leq (1 + \delta/\varepsilon_0)^r. 
$$
Using the binomial theorem, $(1 + \delta/\varepsilon_0)^r - 1 = r \delta/\varepsilon_0 + \sum_{h=2}^{r} {r \choose h} (\delta/\varepsilon_0)^h$. Next, bound ${r \choose h} \leq r^h$ and use the fact that $r \delta/\varepsilon_0 < 1/2$ to conclude that 
$\sum_{h=2}^{r} {r \choose h} (\delta/\varepsilon_0)^h \leq \sum_{h=1}^{\infty}  (r \delta/\varepsilon_0)^h \leq 2 r \delta/\varepsilon_0$. 

On the other hand, using $u_h \geq v_h - \delta$ for all $h$,
$$
\frac{u_1 \ldots u_r}{v_1 \ldots v_r} \geq \prod_{h=1}^r (1 - \delta/v_h) \geq (1 - \delta/\varepsilon_0)^r \geq 1 - r\delta/\varepsilon_0. 
$$
The proof is concluded by observing that
$$
\max \l \{\frac{u_1 \ldots u_r}{v_1 \ldots v_r} - 1, 1 - \frac{u_1 \ldots u_r}{v_1 \ldots v_r} \r\} \leq 2r \delta/\varepsilon_0.
$$

\bibliographystyle{jasa}
\bibliography{cov_refs}

\begin{thebibliography}{43}
\newcommand{\enquote}[1]{``#1''}
\expandafter\ifx\csname natexlab\endcsname\relax\def\natexlab#1{#1}\fi

\bibitem[Agresti(2002)]{agresti2002categorical}
Agresti, A. (2002), \emph{Categorical data analysis}, Vol. 359,
  Wiley-interscience.
\bibitem[Armagan {\normalfont et~al.}(2013{\natexlab{a}})Armagan, Dunson and
  Lee]{armagan2011generalized}
Armagan, A., Dunson, D., and Lee, J. (2013{\natexlab{a}}), \enquote{Generalized
  double Pareto shrinkage,} \emph{Statistica Sinica}, 23, 119--143.
\bibitem[Armagan {\normalfont et~al.}(2013{\natexlab{b}})Armagan, Dunson, Lee,
  Bajwa and Strawn]{armagan2013posterior}
Armagan, A., Dunson, D., Lee, J., Bajwa, W., and Strawn, N.
  (2013{\natexlab{b}}), \enquote{Posterior consistency in high-dimensional
  linear models,} \emph{Biometrika (to appear)}.
\bibitem[Belitser and Ghosal(2003)]{belitser2003adaptive}
Belitser, E., and Ghosal, S. (2003), \enquote{Adaptive Bayesian inference on
  the mean of an infinite-dimensional normal distribution,} \emph{The Annals of
  Statistics}, 31, 536--559.
\bibitem[Bhattacharya and Dunson(2011)]{bhattacharya2011sparse}
Bhattacharya, A., and Dunson, D. (2011), \enquote{Sparse Bayesian infinite
  factor models,} \emph{Biometrika}, 98, 291--306.
\bibitem[Bhattacharya and Dunson(2012)]{anirban2012}
Bhattacharya, A., and Dunson, D. (2012), \enquote{Simplex factor models for
  multivariate unordered categorical data,} \emph{Journal of the American
  Statistical Association}, 107, 362--377.
\bibitem[Bontemps(2011)]{bontemps2011bernstein}
Bontemps, D. (2011), \enquote{Bernstein--von Mises theorems for Gaussian
  regression with increasing number of regressors,} \emph{The Annals of
  Statistics}, 39, 2557--2584.
\bibitem[Bro(1997)]{bro1997}
Bro, R. (1997), \enquote{PARAFAC. Tutorial and applications,}
  \emph{Chemometrics and Intelligent Laboratory Systems}, 38, 149--171.
\bibitem[Candes and Recht(2009)]{candesrecht2009}
Candes, E., and Recht, B. (2009), \enquote{Exact matrix completion via convex
  optimization,} \emph{Foundations of Computational Mathematics}, 9, 717--772.
\bibitem[Carvalho {\normalfont et~al.}(2008)Carvalho, Lucas, Wang, Nevins and
  West]{Carvalho08}
Carvalho, C., Lucas, J., Wang, Q., Nevins, J., and West, M. (2008),
  \enquote{{High-dimensional sparse factor modelling: applications in gene
  expression genomics },} \emph{Journal of the American Statistical
  Association}, 103, 1438--1456.
\bibitem[Carvalho {\normalfont et~al.}(2010)Carvalho, Polson and
  Scott]{carvalho2010horseshoe}
Carvalho, C., Polson, N., and Scott, J. (2010), \enquote{The horseshoe
  estimator for sparse signals,} \emph{Biometrika}, 97, 465--480.
\bibitem[Castillo and van~der Vaart(2012)]{castilloneedles}
Castillo, I., and van~der Vaart, A. (2012), \enquote{Needles and straws in a
  haystack: Posterior concentration for possibly sparse sequences,} \emph{The
  Annals of Statistics}, 40, 2069--2101.
\bibitem[Chartrand(2012)]{chartrand2012}
Chartrand, R. (2012), \enquote{Nonconvex splitting for regularized low-rank
  plus sparse decomposition,} \emph{IEEE Transactions on Signal Processing},
  60, 5810--5819.
\bibitem[Dunson and Xing(2009)]{dunsonxing08}
Dunson, D.~B., and Xing, C. (2009), \enquote{{Nonparametric Bayes modeling of
  multivariate categorical data},} \emph{Journal of the American Statistical
  Association}, 104, 1042--1051.
\bibitem[Fienberg and Rinaldo(2007)]{fienberg2007three}
Fienberg, S., and Rinaldo, A. (2007), \enquote{{Three centuries of categorical
  data analysis: Log-linear models and maximum likelihood estimation},}
  \emph{Journal of Statistical Planning and Inference}, 137, 3430--3445.
\bibitem[Friedlander and Hatz(2005)]{friedlander2005}
Friedlander, M., and Hatz, K. (2005), \enquote{Computing non-negative tensor
  factorizations,} \emph{Optimization Methods and Software}, 23, 631--647.
\bibitem[Ge and Jiang(2006)]{ge2006consistency}
Ge, Y., and Jiang, W. (2006), \enquote{On consistency of Bayesian inference
  with mixtures of logistic regression,} \emph{Neural Computation}, 18,
  224--243.
\bibitem[Gelman {\normalfont et~al.}(2008)Gelman, Jakulin, Pittau and
  Su]{Gelman2008}
Gelman, A., Jakulin, A., Pittau, M., and Su, Y. (2008), \enquote{A weakly
  informative default prior distribution for logistic and other regression
  models,} \emph{Annals of Applied Statistics}, 2, 1360--1383.
\bibitem[Ghosal(1999)]{ghosal1999as}
Ghosal, S. (1999), \enquote{Asymptotic normality of posterior distributions in
  high-dimensional linear models,} \emph{Bernoulli}, 5, 315--331.
\bibitem[Ghosal(2000)]{ghosal2000as}
Ghosal, S. (2000), \enquote{Asymptotic normality of posterior distributions for
  exponential families when the number of parameters tends to infinity,}
  \emph{Journal of Multivariate Analysis}, 74, 49--68.
\bibitem[Ghosal {\normalfont et~al.}(2000)Ghosal, Ghosh and van~der
  Vaart]{ghosal2000convergence}
Ghosal, S., Ghosh, J., and van~der Vaart, A. (2000), \enquote{{Convergence
  rates of posterior distributions},} \emph{Annals of Statistics}, 28,
  500--531.
\bibitem[Hans(2011)]{hans2011elastic}
Hans, C. (2011), \enquote{Elastic net regression modeling with the orthant
  normal prior,} \emph{Journal of the American Statistical Association}, 106,
  1383--1393.
\bibitem[Harshman(1970)]{harshman70}
Harshman, R. (1970), \enquote{{Foundations of the PARAFAC procedure: Models and
  conditions for an ``explanatory'' multi-modal factor analysis},} \emph{UCLA
  Working Papers in Phonetics}, 16, 84.
\bibitem[Jiang(2007)]{jiang2007bayesian}
Jiang, W. (2007), \enquote{Bayesian variable selection for high dimensional
  generalized linear models: convergence rates of the fitted densities,}
  \emph{The Annals of Statistics}, 1487--1511.
\bibitem[Karatzoglou {\normalfont et~al.}(2010)Karatzoglou, Amatriain,
  Baltrunas and Oliver]{Karatzoglou2010}
Karatzoglou, A., Amatriain, X., Baltrunas, L., and Oliver, N. (2010),
  \enquote{Multiverse recommendation: n-dimensional tensor factorization for
  context-aware collaborative filtering,} \emph{Proceedings of the Fourth ACM
  Conference on Recommender Systems}.
\bibitem[Kolda and Bader(2009)]{kolda09tensor}
Kolda, T., and Bader, B. (2009), \enquote{{Tensor decompositions and
  applications},} \emph{SIAM Review}, 51, 455--500.
\bibitem[Lee and Seung(1999)]{lee1999learning}
Lee, D.~D., and Seung, H.~S. (1999), \enquote{Learning the parts of objects by
  non-negative matrix factorization,} \emph{Nature}, 401, 788--791.
\bibitem[Lim and Comon(2009)]{limcomon09}
Lim, L., and Comon, P. (2009), \enquote{{Nonnegative approximations of
  nonnegative tensors},} \emph{Jour. Chemometrics}, 432--441.
\bibitem[Liu {\normalfont et~al.}(2012)Liu, Liu, Wonka and Ye]{liu2012}
Liu, J., Liu, J., Wonka, P., and Ye, J. (2012), \enquote{Sparse non-negative
  tensor factorization using columnwise coordinate descent,} \emph{Pattern
  Recognition}, 45, 649--656.
\bibitem[Lucas {\normalfont et~al.}(2006)Lucas, Carvalho, Wang, Bild, Nevins
  and West]{Lucasetal06}
Lucas, J.~E., Carvalho, C., Wang, Q., Bild, A., Nevins, J., and West, M.
  (2006), \enquote{Sparse statistical modelling in gene expression genomics,}
  in \emph{Bayesian Inference for Gene Expression and Proteomics}, eds. K.~Do,
  P.~M\"uller and M.~Vannucci, Cambridge University Press, pp.\  155--176.
\bibitem[Paatero and Tapper(1994)]{paatero1994positive}
Paatero, P., and Tapper, U. (1994), \enquote{Positive matrix factorization: A
  non-negative factor model with optimal utilization of error estimates of data
  values,} \emph{Environmetrics}, 5, 111--126.
\bibitem[Park and Casella(2008)]{park2008bayesian}
Park, T., and Casella, G. (2008), \enquote{The Bayesian lasso,} \emph{Journal
  of the American Statistical Association}, 103, 681--686.
\bibitem[Pati {\normalfont et~al.}(2013{\natexlab{a}})Pati, Bhattacharya,
  Pillai and Dunson]{debdeep2013}
Pati, D., Bhattacharya, A., Pillai, N., and Dunson, D. (2013{\natexlab{a}}),
  \enquote{Posterior contraction in sparse {B}ayesian factor models for massive
  covariance matrices,} \emph{arXiv:1206.3627}.
\bibitem[Pati {\normalfont et~al.}(2013{\natexlab{b}})Pati, Dunson and
  Tokdar]{pati2013jmva}
Pati, D., Dunson, D.~B., and Tokdar, S.~T. (2013{\natexlab{b}}),
  \enquote{Posterior consistency in conditional distribution estimation,}
  \emph{Journal of Multivariate Analysis}.
\bibitem[Polson and Scott(2010)]{polson2010shrink}
Polson, N., and Scott, J. (2010), \enquote{{Shrink globally, act locally:
  Sparse Bayesian regularization and prediction},} in \emph{Bayesian Statistics
  9 (J.M. Bernardo, M.J. Bayarri, J.O. Berger, A.P. Dawid, D. Heckerman, A.F.M.
  Smith and M. West, eds.)}, Oxford University Press, New York, pp.\  501--538.
\bibitem[Scott and Berger(2010)]{scott2010bayes}
Scott, J., and Berger, J. (2010), \enquote{Bayes and empirical-Bayes
  multiplicity adjustment in the variable-selection problem,} \emph{The Annals
  of Statistics}, 38, 2587--2619.
\bibitem[Sethuraman(1994)]{sethuraman1994constructive}
Sethuraman, J. (1994), \enquote{{A constructive definition of Dirichlet
  priors},} \emph{Statistica Sinica}, 4, 639--650.
\bibitem[Shen {\normalfont et~al.}(2011)Shen, Tokdar and
  Ghosal]{shen2011adaptive}
Shen, W., Tokdar, S., and Ghosal, S. (2011), \enquote{Adaptive Bayesian
  multivariate density estimation with Dirichlet mixtures,} \emph{arXiv
  preprint arXiv:1109.6406}.
\bibitem[Talagrand(1996)]{talagrand1996new}
Talagrand, M. (1996), \enquote{A new look at independence,} \emph{The Annals of
  Probability}, 24, 1--34.
\bibitem[van~der Vaart and van Zanten(2008)]{van2008rates}
van~der Vaart, A., and van Zanten, J. (2008), \enquote{Rates of contraction of
  posterior distributions based on Gaussian process priors,} \emph{The Annals
  of Statistics}, 36, 1435--1463.
\bibitem[Vershynin(2010)]{vershynin2010introduction}
Vershynin, R. (2010), \enquote{Introduction to the non-asymptotic analysis of
  random matrices,} \emph{Arxiv preprint arxiv:1011.3027}.
\bibitem[West(2003)]{West03}
West, M. (2003), \enquote{{Bayesian factor regression models in the ``large
  \emph{p}, small \emph{n}'' paradigm},} in \emph{Bayesian Statistics 7 (J.M.
  Bernardo, M.J. Bayarri, J.O. Berger, A.P. Dawid, D. Heckerman, A.F.M. Smith
  and M. West, eds.)}, Oxford University Press, New York, pp.\  733--742.
\bibitem[Yang and Dunson(2013)]{yang2013bayesian}
Yang, Y., and Dunson, D.~B. (2013), \enquote{Bayesian conditional tensor
  factorizations for high-dimensional classification,} \emph{arXiv preprint
  arXiv:1301.4950}.
\end{thebibliography}
\end{document}